\documentclass{emulateapj}

\begin{document}

\shortauthors{Luhman et al.}
\shorttitle{Survey for Members of Taurus with SDSS}

\title{A Survey for New Members of the Taurus Star-Forming Region with
the Sloan Digital Sky Survey\altaffilmark{1}}

\author{
K. L. Luhman\altaffilmark{2,3},
E. E. Mamajek\altaffilmark{4},
S. J. Shukla\altaffilmark{5},
and N. P. Loutrel\altaffilmark{6}
}

\altaffiltext{1}
{Based on observations performed with the Sloan Digital Sky Survey,
Hobby-Eberly Telescope, NASA Infrared Telescope Facility, 
Gemini Observatory, and Canada-France-Hawaii Telescope.}

\altaffiltext{2}{Department of Astronomy and Astrophysics, The Pennsylvania
State University, University Park, PA 16802; kluhman@astro.psu.edu}

\altaffiltext{3}{Center for Exoplanets and Habitable Worlds, The
Pennsylvania State University, University Park, PA 16802, USA}

\altaffiltext{4}{Department of Physics and Astronomy, The University of
Rochester, Rochester, NY 14627}

\altaffiltext{5}{Institute of Astronomy, Madingley Road, Cambridge CB3 0HA, UK}

\altaffiltext{6}{eXtreme Gravity Institute, Department of Physics, Montana State
University, Bozeman, MT 59715}

\begin{abstract}

Previous studies have found that $\sim1$~deg$^2$ fields surrounding the
stellar aggregates in the Taurus star-forming region exhibit a surplus of
solar-mass stars relative to denser clusters like IC~348 and the Orion
Nebula Cluster. To test whether this difference reflects mass segregation
in Taurus or a variation in the IMF, we have performed a survey for members
of Taurus across a large field ($\sim$40~deg$^2$) that was imaged by the
Sloan Digital Sky Survey (SDSS). We obtained optical and near-infrared spectra
of candidate members identified with those images and the Two Micron All Sky
Survey, as well as miscellaneous candidates that were selected with several
other diagnostics of membership. We have classified 22 of the candidates as
new members of Taurus, which includes one of the coolest known members
(M9.75). Our updated census of members within the SDSS field shows
a surplus of solar-mass stars relative to clusters, although it
is less pronounced than in the smaller fields towards the stellar aggregates
that were surveyed for previously measured mass functions in Taurus.
In addition to spectra of our new members, we include
in our study near-IR spectra of roughly half of the known members of Taurus,
which are used to refine their spectral types and extinctions. We also
present an updated set of near-IR standard spectra for classifying young
stars and brown dwarfs at M and L types.

\end{abstract}


\section{Introduction}
\label{sec:intro}

The Taurus complex of dark clouds is one of the nearest star-forming regions
\citep[$d=140$~pc,][]{wic98,loi05,tor07,tor09,tor12} and 
is relatively well-populated with $\sim400$ known members.
Many of its members reside in loose aggregates near
the dark clouds while others are scattered more widely across the
$\sim100$~deg$^2$ extent of the cloud complex. 
Even in the aggregates, the stellar densities are only $\sim10$~pc$^{-3}$,
which is 100--1000 times lower than the densities of the
most compact clusters in other nearby star-forming regions.
Because the stars and cloud cores in Taurus are so sparsely distributed,
they have served as some of the best available targets for studying the
formation of stars in relative isolation.
The close proximity and low density of Taurus also made it amenable to
observations with early telescopes at X-ray, infrared (IR), and radio
wavelengths that had low sensitivity and resolution. As a result,  
much of the foundational observational work in the field of star formation
has been performed in Taurus \citep{ken08}.

Stars and brown dwarfs within Taurus have been sought with a variety
of diagnostics of youth and membership, consisting of 
photometric variability \citep{joy45,joy49},
emission lines \citep{fei83,her86,bri93,bri99},
proper motions \citep{jon79,har91,gom92}, UV emission \citep{fin10,gom15},
optical and near-infrared (IR) photometry
\citep{lr98,bri98,bri02,mar01,luh03tau,luh00,luh04tau,luh06tau1,gui06,sle06},
X-ray emission 
\citep{fei81,wal81,fei87,wal88,str94,neu95,car96,wic96,bri97,bri99,gud07,sce07,sce08},
and mid-IR emission 
\citep{bei86,bei92,har88,mye87,ken90,ken94,reb10,reb11,luh06tau2,luh09fu,luh09tau,moo13,esp14}.
Each survey has been capable of finding members that have specific
ranges of mass, extinction, location, and evolutionary stage.
The current census of members has the highest level of completeness
in $\sim1$~deg$^2$ fields surrounding the richest stellar aggregates.
The initial mass function (IMF) constructed from the known members
within those fields \citep{bri02,luh04tau,luh09tau} exhibits a large surplus
of solar-mass stars relative to the mass functions of clusters like the
Orion Nebula Cluster, IC~348, and Chamaeleon~I
\citep{hil97,hc00,mue02,mue03,luh03ic,luh07cha}.
Given that more massive stars are often found preferentially near the
centers of star-forming clusters \citep[e.g.,][]{hil98,mue03}, it is
plausible that the anomalous nature of the IMFs measured in the Taurus
aggregates is at least partially due to mass segregation. 

To better determine the degree to which the IMF in Taurus differs from
mass functions in other nearby star-forming regions, we have performed
a survey for new members of Taurus across a large area of the region
($\sim40$~deg$^2$) using optical images from the Sloan Digital Sky
Survey \cite[SDSS;][]{yor00,fin04} in conjunction with near-IR data from the
Two Micron All Sky Survey \citep[2MASS,][]{skr06}.
In our presentation of this survey, we begin by compiling a list of all known
members of Taurus from previous studies (Section~\ref{sec:previous}).
We then select candidate members based on their positions in color-magnitude
diagrams constructed from SDSS and 2MASS data and several other diagnostics
of membership (Section~\ref{sec:select}) and use optical and near-IR
spectra to measure their spectral types and assess their membership
(Section~\ref{sec:spectra}). We conclude by using our updated census of
members to check for IMF variations between the SDSS fields, the smaller
regions surrounding the stellar aggregates in Taurus, and denser clusters
like IC~348 (Section~\ref{sec:imf}).

\section{Census from Previous Studies}
\label{sec:previous}

Before presenting our survey for new members of Taurus, we describe
the list of previously known members that we have adopted. We began with the
census compiled by \citet{luh09tau,luh10tau}, which consisted of 352
members that are resolved by 2MASS or the Infrared Array Camera 
\citep[IRAC;][]{faz04} on the {\it Spitzer Space Telescope} \citep{wer04}.
The components of a given multiple system that are unresolved in those images
appear as a single entry.
We now reject two of the stars from that catalog, HBC~372 and HBC~407.
HBC~372 has been previously described as a relatively old, Li-depleted member
of Taurus \citep{ses08}. However, it is sufficiently faint relative to
other Taurus members near its spectral type ($\sim3.4$~mag below the 
median sequence) that it is not plausibly related to the star-forming event
that produced the known members. We note that an edge-on disk cannot
explain the faint photometry for this star since it does not exhibit
evidence of a disk in mid-IR photometry \citep{luh10tau,esp14}.
HBC~407 is rejected because its proper motion
\citep[e.g.,][]{ber06,duc05,roe10,zac13} differs from those of the known
members \citep{luh09tau}.

To the list of members from \citet{luh09tau,luh10tau}, we
have added the 25 new members from \citet{esp14}, the three young
stars with proper motions consistent with membership from \citet{fin10},
and four new members from \citet{reb10}, consisting of
2MASS J04251550+2829275, 2MASS J04355760+2253574, 2MASS J04355949+2238291,
and 2MASS J04380191+2519266.
\citet{esp14} described evidence of membership from previous studies
for HD~285957, V1195 Tau, HD~31305, HD~286178, and RXJ 0432.7+1809, which
are included in our census.
We also adopt the following stars as members:
LH~0419+15 based on Li absorption and gravity-sensitive spectral
features \citep[][this work]{rei99}\footnote{\citet{esp14} reported
that LH~0419+15 exhibited excess emission at 12~\micron\ from the
{\it Wide-field Infrared Survey Explorer} \citep[{\it WISE},][]{wri10},
but the detection in that band appears to be spurious based on closer
inspection of those images.},
2MASS J04242321+2650084 based on IR excess emission and proper motion
\citep{reb10,cie12,zac13}, RXJ 0432.8+1735 based on Li absorption, IR excess
emission, and proper motion \citep{car96,mar99,pad06}, and
SST Tau 041831.2+282617 based on mid-IR spectral features \citep{fur11}.
We arrive 391 known members of Taurus from previous studies
that are resolved by 2MASS or IRAC. In this work, we have
identified 22 additional members, resulting in a total of 413 known members.
We present the full sample of members in Table~\ref{tab:mem}.

In a multiplicity survey of Taurus members, \citet{dae15}
included an ``extended sample" of G-K stars that was described as
co-moving with the traditional members and associated with the Taurus clouds,
but was relatively old based on weak Li absorption ($\sim$20~Myr).
We find that most of the stars in that extended sample
have proper motions or radial velocities that are inconsistent with a physical
association with the traditional members \citep[e.g.,][]{wic00,roe10,zac13}.
In addition, the Li strengths of most of those stars are consistent with
ages of $\sim100$~Myr based on a comparison to the Pleiades \citep{sod93}.
It is possible that a few stars in the extended sample from \citet{dae15} are
associated with Taurus, but none show convincing evidence of membership,
and the sample is likely dominated by young main sequence stars that are
unrelated to Taurus \citep{bri97}.

\section{Identification of Candidate Members}
\label{sec:select}

\begin{figure}
\epsscale{1}
\plotone{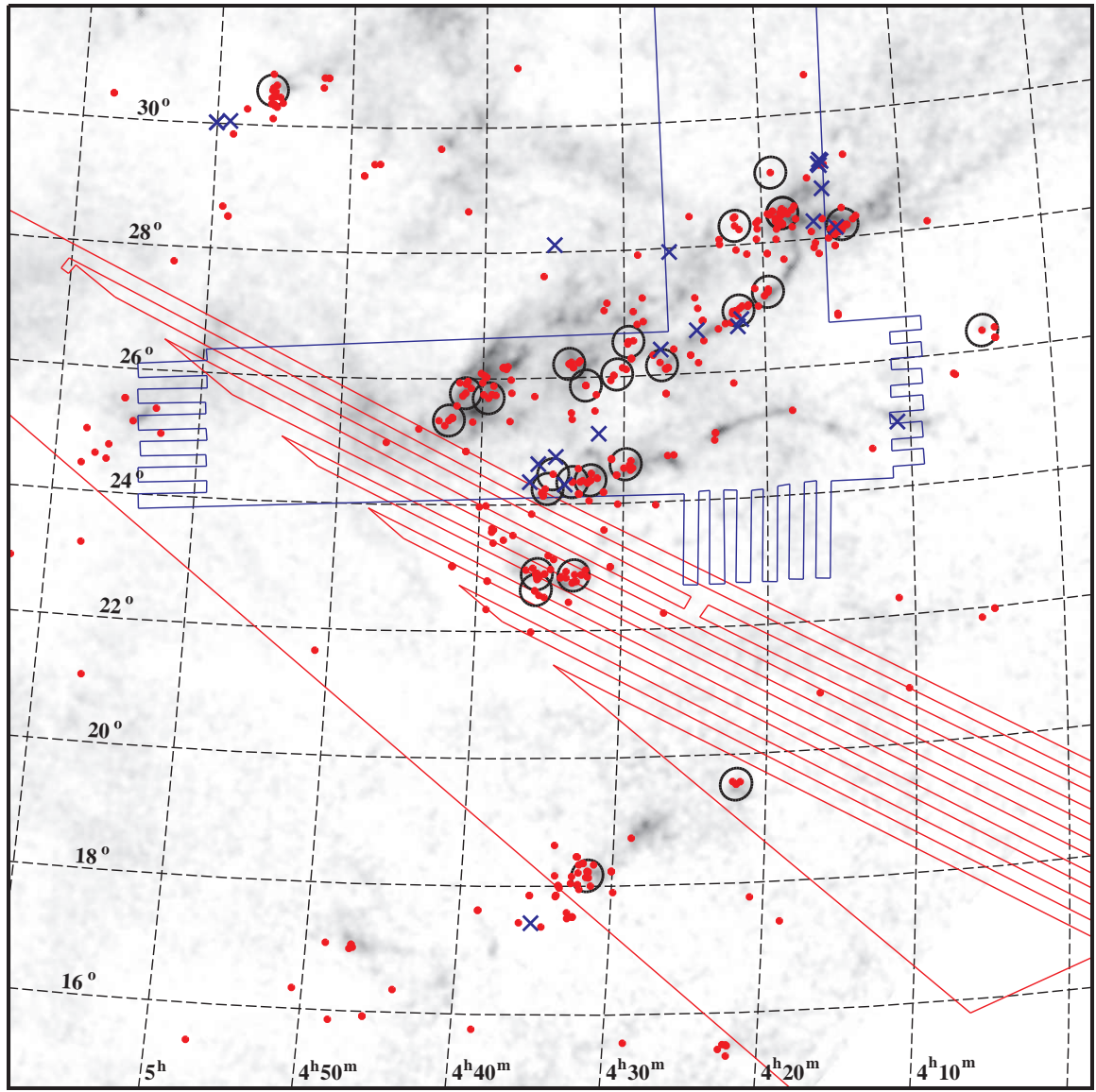}
\caption{
Spatial distribution of previously known members of the Taurus star-forming
region (filled circles) and new members from this work (crosses).
The field enclosed by the blue lines was imaged by SDSS in \citet{fin04}.
Most of the new members were identified as candidates with photometry from SDSS
and 2MASS in that field. After the observations in \citet{fin04}, SDSS
observed additional areas that are outlined with the red lines.
The fields imaged with {\it XMM-Newton} through the XEST program
are indicated \citep[large circles,][]{gud07}.
The dark clouds in Taurus are displayed with a map of extinction
\citep[gray scale,][]{dob05}.
}
\label{fig:map}
\end{figure}

For the purposes of commissioning, testing, and calibration, SDSS obtained
images of areas that were outside of the official survey region \citep{fin04}.
Portions of the Taurus star-forming region were included in those extra
observations.
In Figure~\ref{fig:map}, the SDSS fields in Taurus are indicated in a map
of the dark clouds and known members.
\citet{fin04} reported imaging of the field outlined in blue,
which covers $\sim40$~deg$^2$. The area marked in red was
observed by SDSS after that study. We have searched for members of Taurus
in the field from \citet{fin04}, and have not utilized the newer SDSS data.

The SDSS observations described by \citet{fin04} were not included
in the early data releases for SDSS, and instead were processed and
disseminated separately in that study. All of the SDSS data in Taurus have been
part of the official data releases starting with Data Release 8
\citep{aih11}. The version of the {\rm photo} pipeline in those later
releases was v5\_6\_3, which was updated somewhat from version v5\_4\_25 used
in \citet{fin04}. The algorithmic changes between these reductions were
small and the resulting photometric measurements were unchanged at the
$\sim$2\% level.
For our study, we used the original version of the data that was presented by
\citet{fin04}. In their catalog of sources detected by SDSS, \citet{fin04}
reported photometry for a number of aperture sizes. We have used their
magnitudes for an aperture radius of $1\farcs745$. 
The SDSS images were obtained in
the $u$, $g$, $r$, $i$, and $z$ filters. Because we are searching for
stars that are cool and are reddened by dust extinction, the latter two bands
offer the greatest sensitivity for our survey. For each SDSS source,
we searched for a counterpart within $1\arcsec$ in the 2MASS Point Source
Catalog, which provides photometry in $J$, $H$, and $K_s$.
Approximately 2 million sources were detected in both SDSS and 2MASS
in the field from \citet{fin04}.

\begin{figure}
\epsscale{1.1}
\plotone{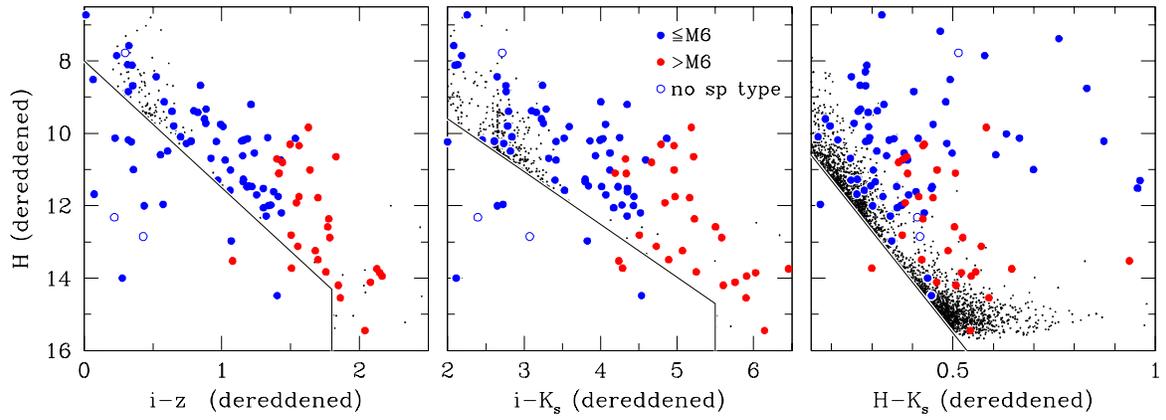}
\caption{
Extinction-corrected CMDs from SDSS and 2MASS for the known members of Taurus
(large filled circles) and other sources (small points) that are within the
SDSS field from \citet{fin04} (Figure~\ref{fig:map}).
Candidate members have been selected based on positions above the
solid boundaries. Stars below the boundaries have been omitted for clarity.
}
\label{fig:cmd}
\end{figure}

To identify candidate members of Taurus with SDSS and 2MASS,
we constructed extinction-corrected diagrams of $i-z$, $i-K_s$, and $H-K_s$
versus $H$, which are shown in Figure~\ref{fig:cmd}. The extinctions for
sources in these diagrams were estimated in the manner described by
\citet{luh03tau}.
In each color-magnitude diagram (CMD), we have marked a boundary along
the lower envelope of the sequence formed by most known members for use in
selecting candidate members. We have omitted the field stars appearing
below the boundaries so that the known members in those areas of the
diagrams can be more easily seen.
Several known members are unusually faint for their colors and
fall below the boundaries in the CMDs, which is normally attributed
to scattered light dominating the observed flux.
Some of those members are known to have edge-on disks or protostellar
envelopes, and most of the others have been suspected of harboring
such structures based on similar CMDs from previous studies.
We identified sources as candidate members if they appeared above the
boundaries in both the $i-K_s$ and $H-K_s$ CMDs, were not rejected in
the $i-z$ CMD, were not rejected as non-members based on proper motions
\citep{roe10,zac13}, and exhibited $J-H>0.5$ ($\gtrsim$K4) after correction
for extinction. We describe spectroscopy of the resulting candidates in
the next section.

In addition to the candidates from the CMDs in the SDSS field, we have
selected for spectroscopy a sample of miscellaneous candidates
that were identified with several other methods.
One of these candidates, 2MASS J04144158+2809583, was selected based on
its colors in images from IRAC on {\it Spitzer}
and WIRCam on the Canada-France-Hawaii Telescope (CFHT). The WIRCam images
were taken in the $J$, $H$, and $K_s$ filters and cover 0.8~deg$^2$ of the
B209 cloud, which contains the westernmost rich aggregate in Taurus.
They were obtained through program 05BF35 (J.-L. Monin) and are publicly
available in the CFHT archive. We measured photometry for all sources
detected in those images and matched them to objects detected by
IRAC \citep{luh06tau2,luh10tau}. To search for low-mass brown dwarfs
($>$M8) in those data, we applied the following criteria:
not detected in optical images form the Digitized Sky Survey,
not resolved as a galaxy in the WIRCam images, 
not previously classified as a member or non-member,
$H<17$, $H-K_s>0.6$, $K_s-[4.5]>0.7$ where [4.5] is the 4.5~\micron\ band
of IRAC, $H-K_s>J-H-0.7/1.55+0.4$, 
and photometric errors less than 0.1~mag in $J$, $H$, $K_s$, and [4.5].
These criteria produced one candidate, 2MASS J04144158+2809583.
We also selected for spectroscopy several objects from the IRAC surveys
of Taurus that are candidate disk-bearing stars based on their red mid-IR
colors \citep{luh06tau2,luh10tau} and three stars from previous studies
that have evidence of membership from proper motions \citep{roe10,zac13}
and signatures of youth in the form of H$\alpha$ emission \citep{koh99},
X-ray emission, or red mid-IR colors from {\it WISE}.
Finally, we obtained spectra of a candidate companion to IRAS 04125+2902
identified by \citet{luh09tau} and a candidate companion to GZ Aur that
was noticed during our spectroscopy of that star.

\section{Spectroscopy}
\label{sec:spectra}

\subsection{Observations}
\label{sec:obs}

We have obtained optical and near-IR spectra of 40 candidate members of Taurus
that were selected in the previous section and possible companions to
two of these candidates.
We also observed 104 known members to improve the measurements
of their spectral types and extinctions.
The spectra were collected with the Gemini Near-Infrared Imager
\citep[NIRI,][]{hod03} using the $K$-band grism and $0\farcs47$ slit
(1.9--2.5~\micron, $R=700$), 
the Gemini Multi-Object Spectrograph \citep[GMOS,][]{hoo04} using the
400 l~mm$^{-1}$ grating and $0\farcs75$ slit (0.56--1~\micron, $R=1500$),
the Marcario Low-Resolution Spectrograph (LRS) on the Hobby-Eberly Telescope
(HET) using the G3 grism and $2\arcsec$ slit (0.63--0.91~\micron, $R=1100$),
and SpeX \citep{ray03} at the NASA Infrared Telescope Facility (IRTF)
using either the prism or SXD mode ($R=150/750$) and $0\farcs8$ slit 
(0.8--2.5~\micron).
The SpeX data were reduced with the Spextool package \citep{cus04} and
corrected for telluric absorption in the manner described by \citet{vac03}.
The spectra from the other instruments were reduced with similar methods
using routines within IRAF.

In the next section, we classify 22 and 18 of the candidates in
our spectroscopic sample as members and non-members, respectively.
Two of these members have candidate companions that were also observed
spectroscopically, both of which have uncertain membership based on our spectra.
The members and non-members are listed in Tables~\ref{tab:new} and
\ref{tab:non}, respectively, which provide the spectral classifications,
the methods of selection, and the instruments and dates for the spectroscopy.
The dates of the SpeX observations are also indicated for both new and
previously known members in Table~\ref{tab:mem}.

\begin{figure}
\epsscale{1}
\plotone{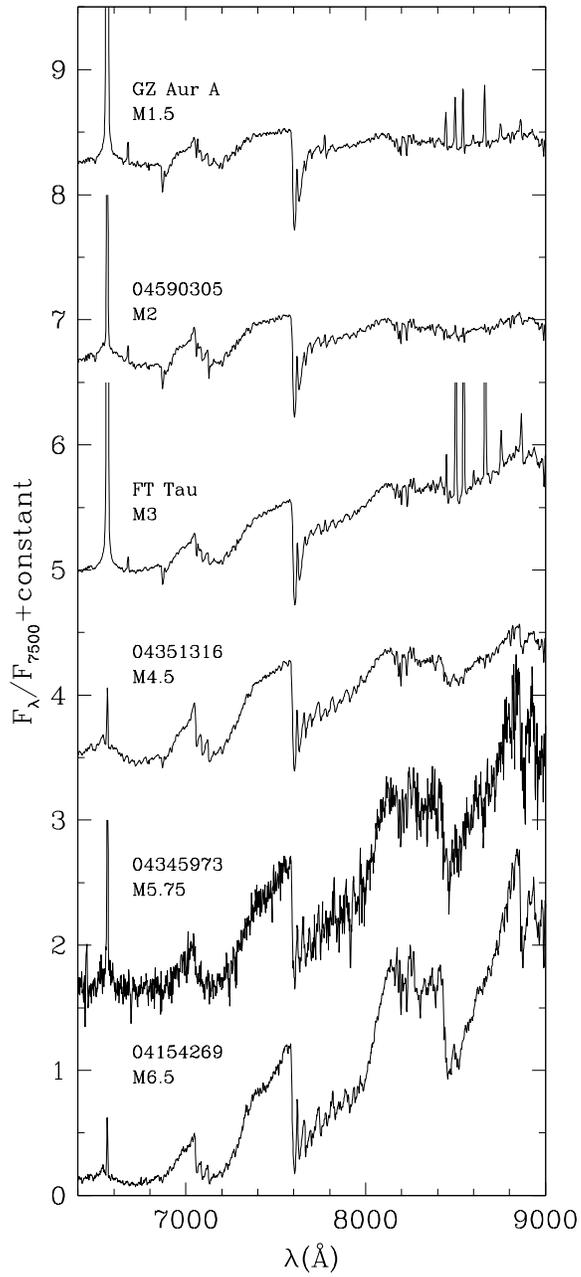}
\caption{
Optical spectra of new and previously known members of Taurus. The spectral
types measured from these data are indicated. The spectra have a resolution
of 7~\AA. The data used to create this figure are available.
}
\label{fig:op}
\end{figure}

\begin{figure}
\epsscale{1}
\plotone{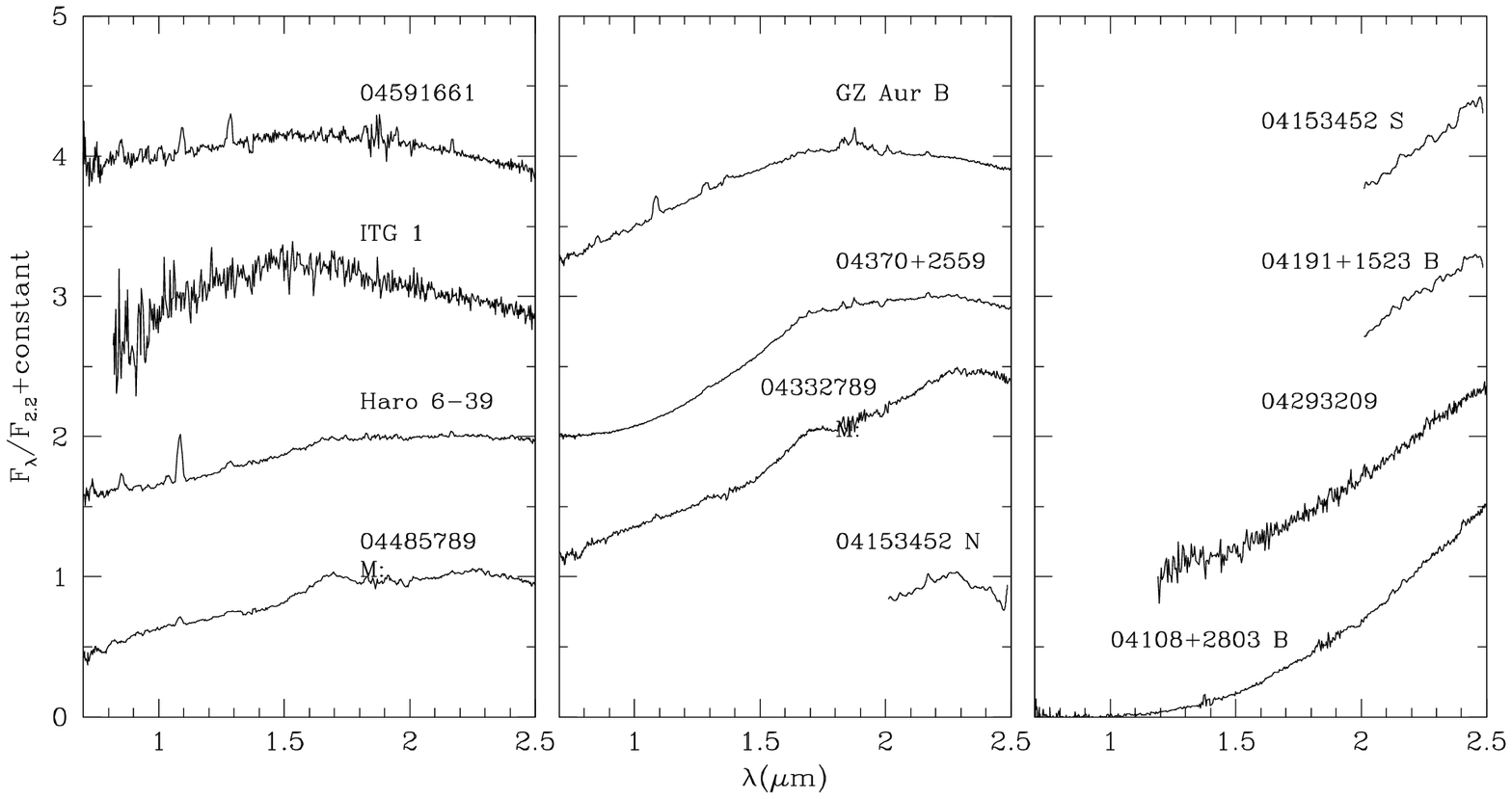}
\caption{
Near-IR spectra of members of Taurus that lack measured spectral types
\citep[][this work]{luh06tau2,luh09tau,esp14}.
These data have a resolution of $R=150$.
The data used to create this figure are available.
}
\label{fig:ir1}
\end{figure}

\begin{figure}
\epsscale{1}
\plotone{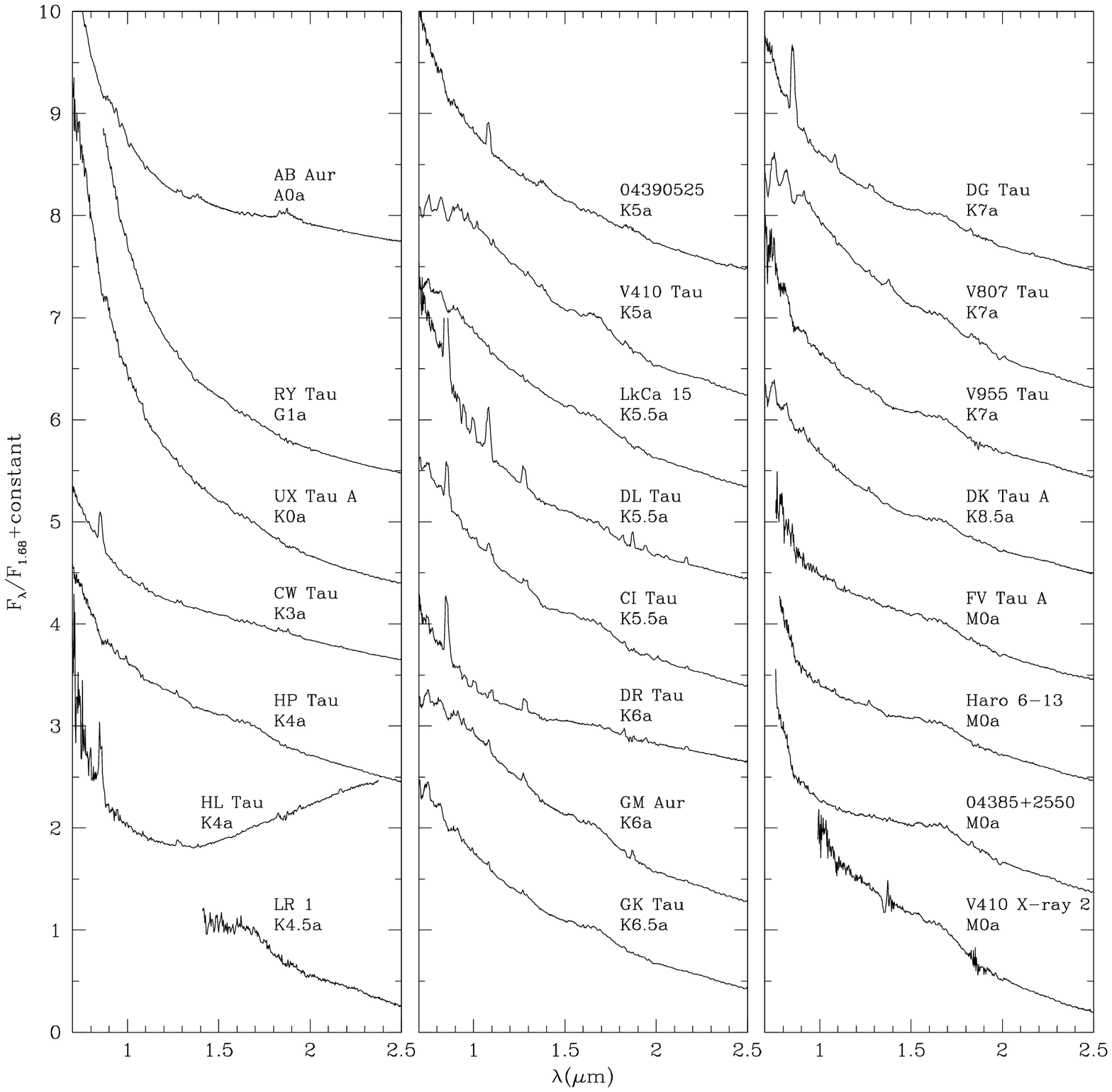}
\caption{
Near-IR spectra of members of Taurus 
\citep[][this work]{luh06tau1,luh06tau2,luh07edgeon,luh09fu,luh09tau,mue07,esp14}.
The spectral types denoted with ``a" have been adopted from optical spectra
because accurate types could not be measured from these IR data
or the objects serve as standards for classifying our IR spectra
(see Appendix). The remaining types have been measured from these
spectra. The spectra have been dereddened to match the slopes of
standards near 1~\micron. These data have a resolution of $R=150$.
The data used to create this figure are available.
}
\label{fig:ir2}
\end{figure}

\begin{figure}
\epsscale{1}
\plotone{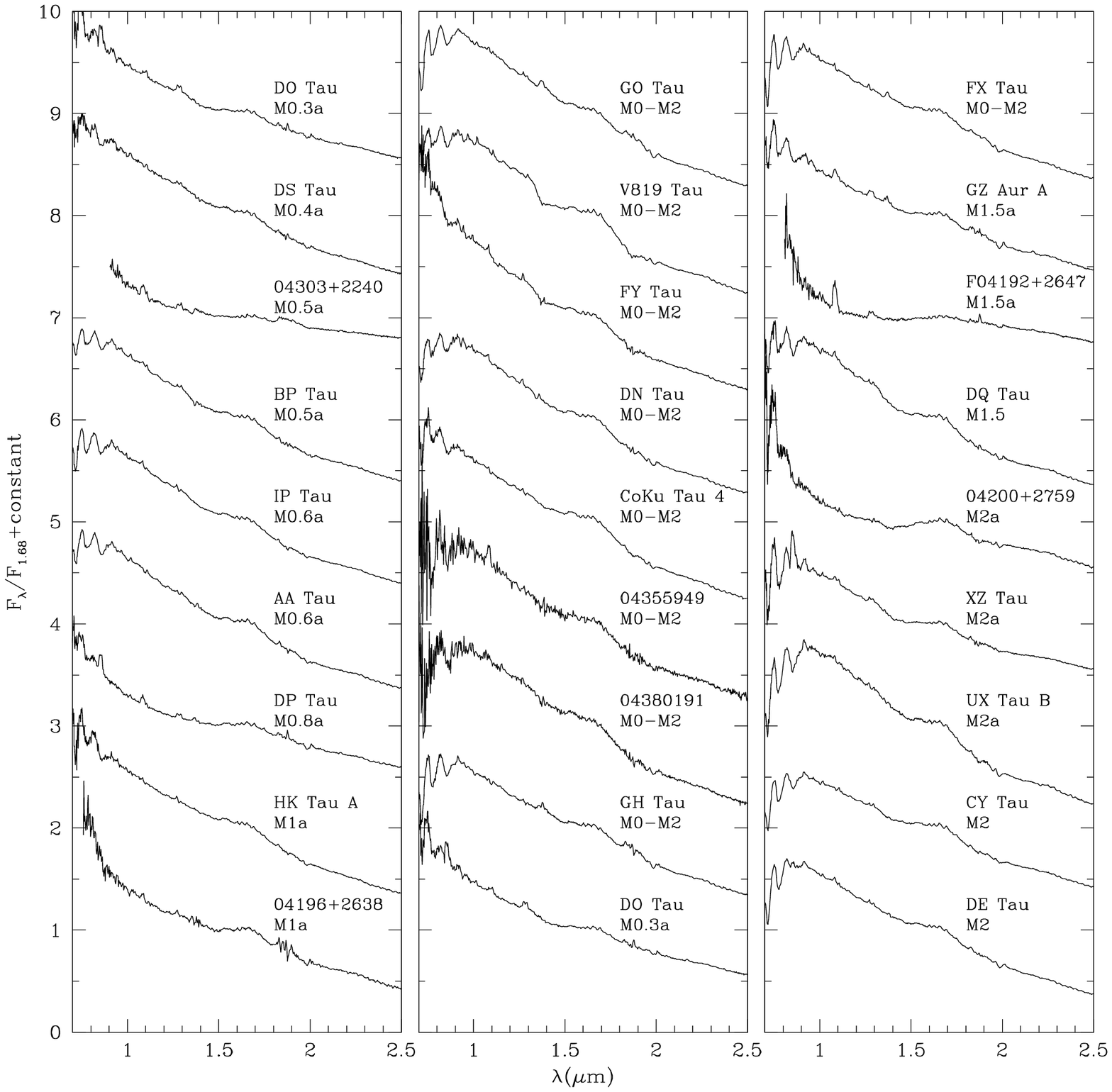}
\caption{
More near-IR spectra of members of Taurus (see Figure~\ref{fig:ir2}).
The data used to create this figure are available.
}
\label{fig:ir3}
\end{figure}

\begin{figure}
\epsscale{1}
\plotone{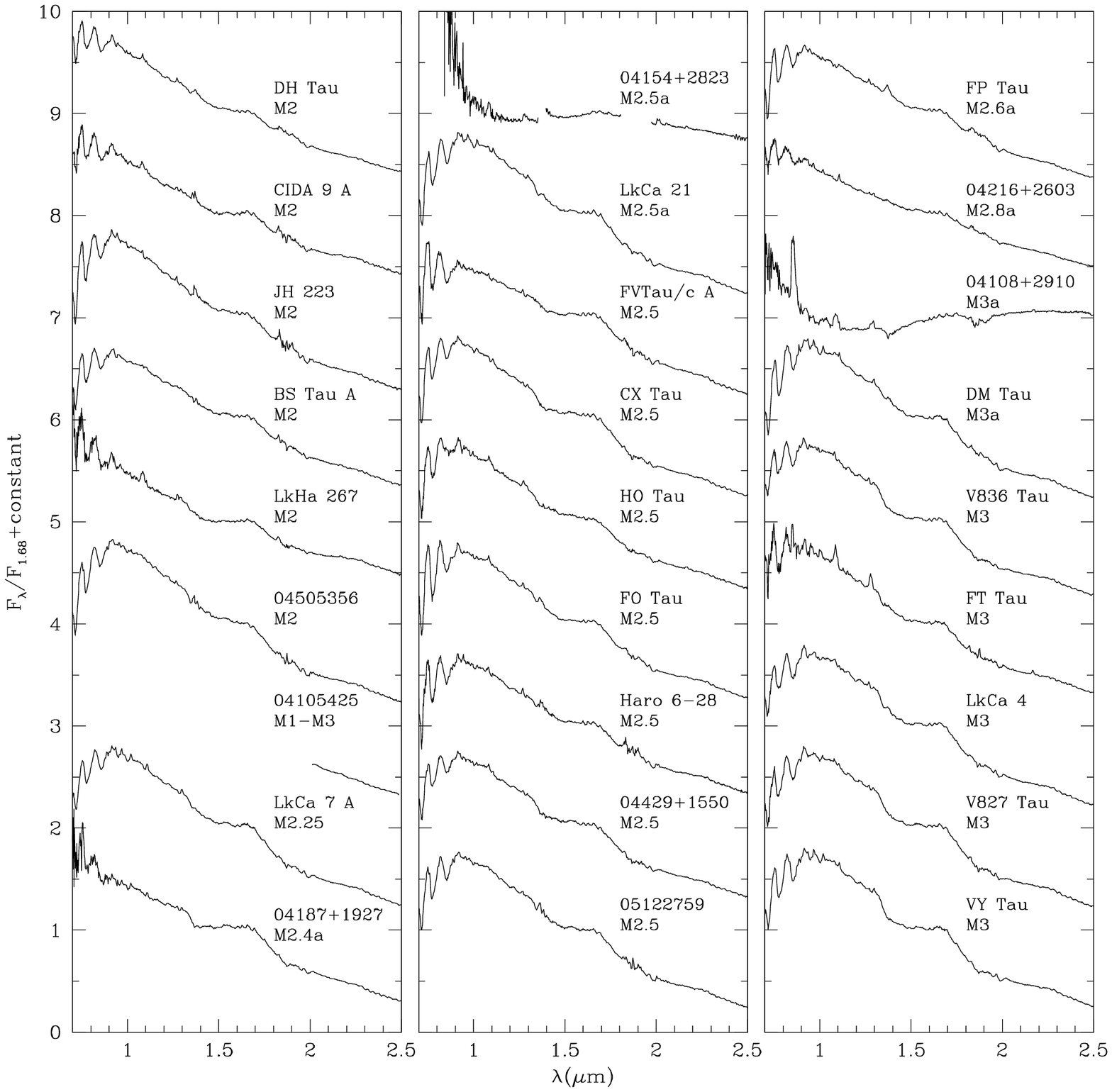}
\caption{
More near-IR spectra of members of Taurus (see Figure~\ref{fig:ir2}).
The data used to create this figure are available.
}
\label{fig:ir4}
\end{figure}

\begin{figure}
\epsscale{1}
\plotone{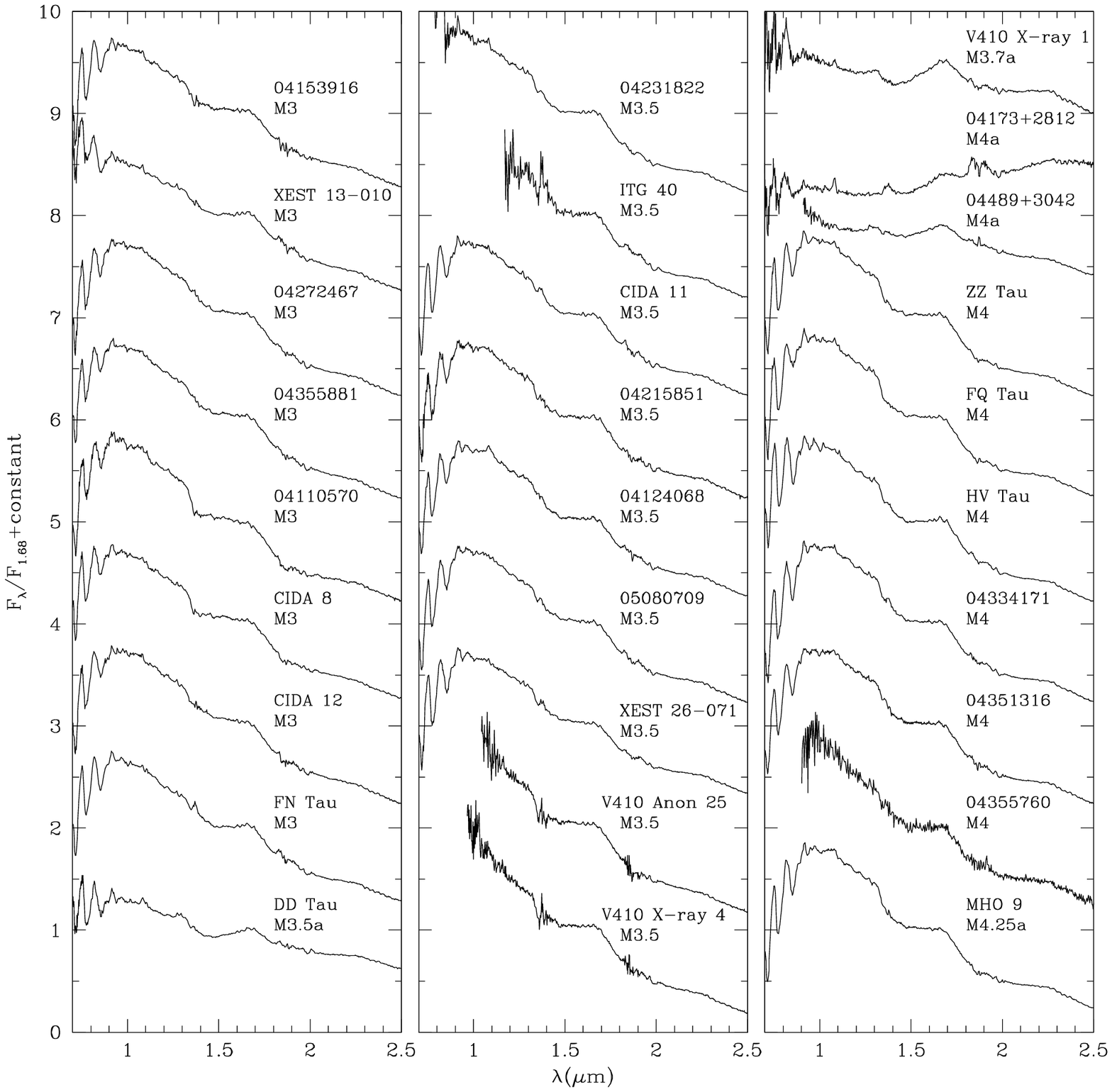}
\caption{
More near-IR spectra of members of Taurus (see Figure~\ref{fig:ir2}).
The data used to create this figure are available.
}
\label{fig:ir5}
\end{figure}

\begin{figure}
\epsscale{1}
\plotone{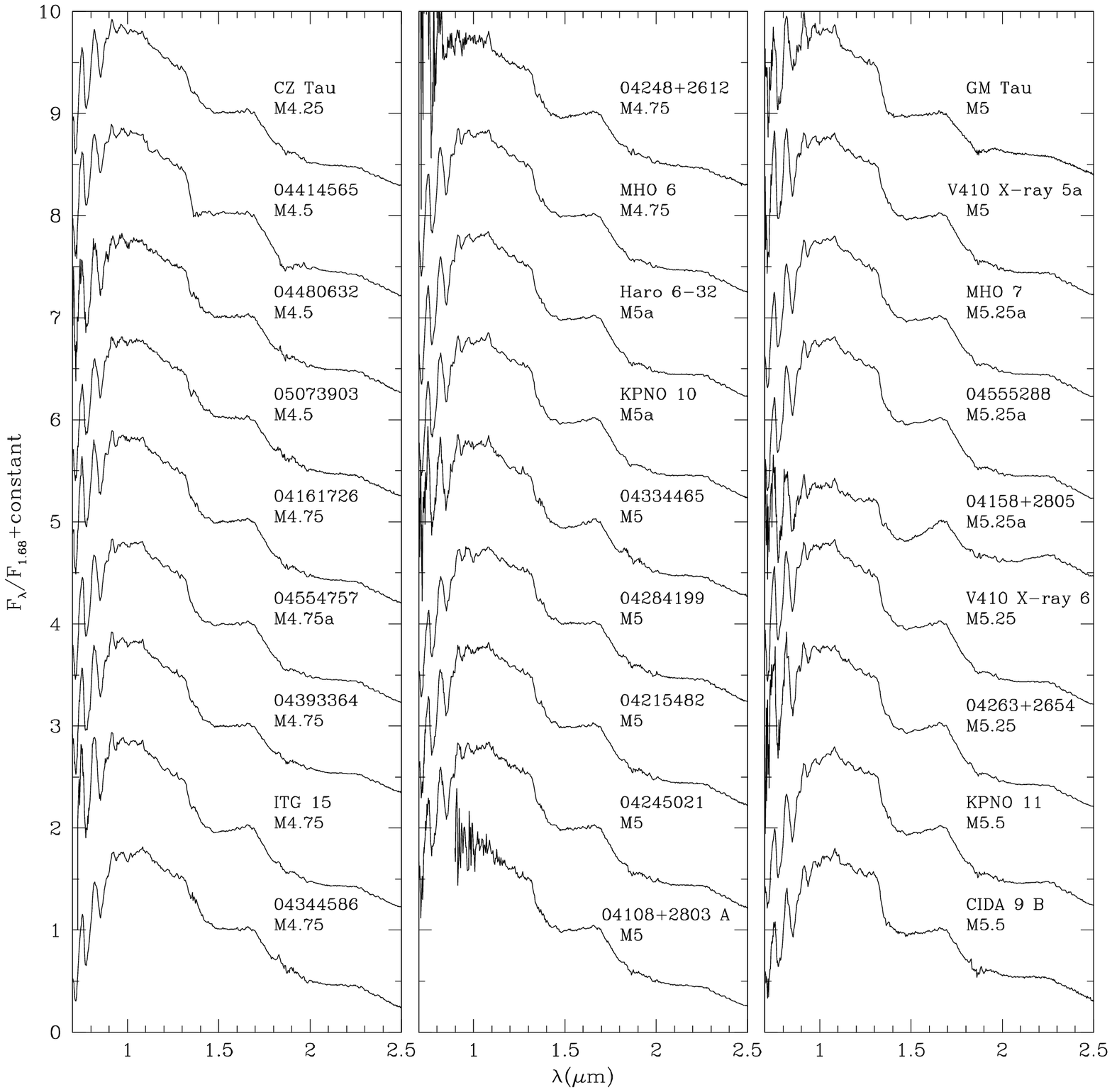}
\caption{
More near-IR spectra of members of Taurus (see Figure~\ref{fig:ir2}).
The data used to create this figure are available.
}
\label{fig:ir6}
\end{figure}

\begin{figure}
\epsscale{1}
\plotone{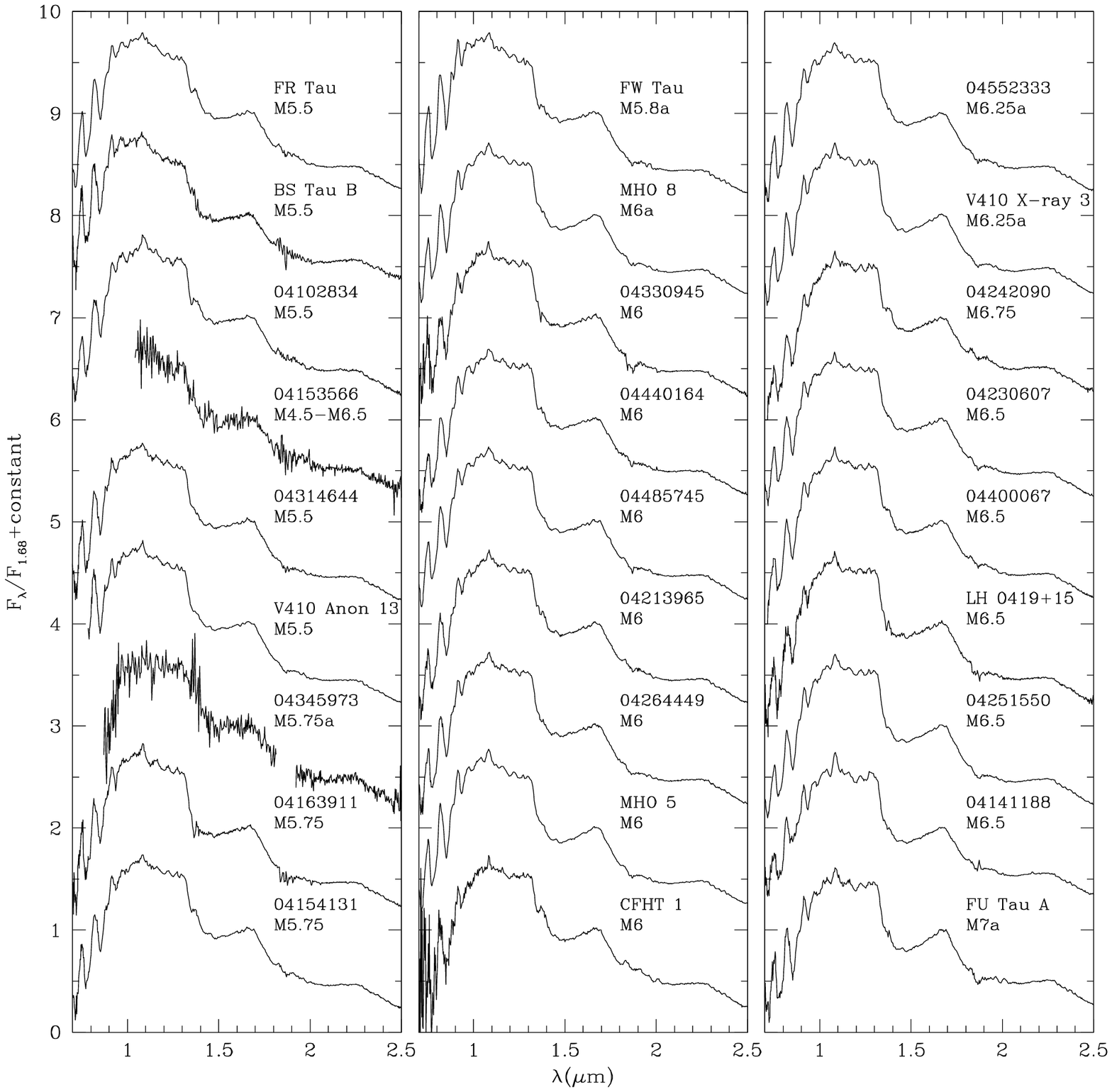}
\caption{
More near-IR spectra of members of Taurus (see Figure~\ref{fig:ir2}).
The data used to create this figure are available.
}
\label{fig:ir7}
\end{figure}

\begin{figure}
\epsscale{1}
\plotone{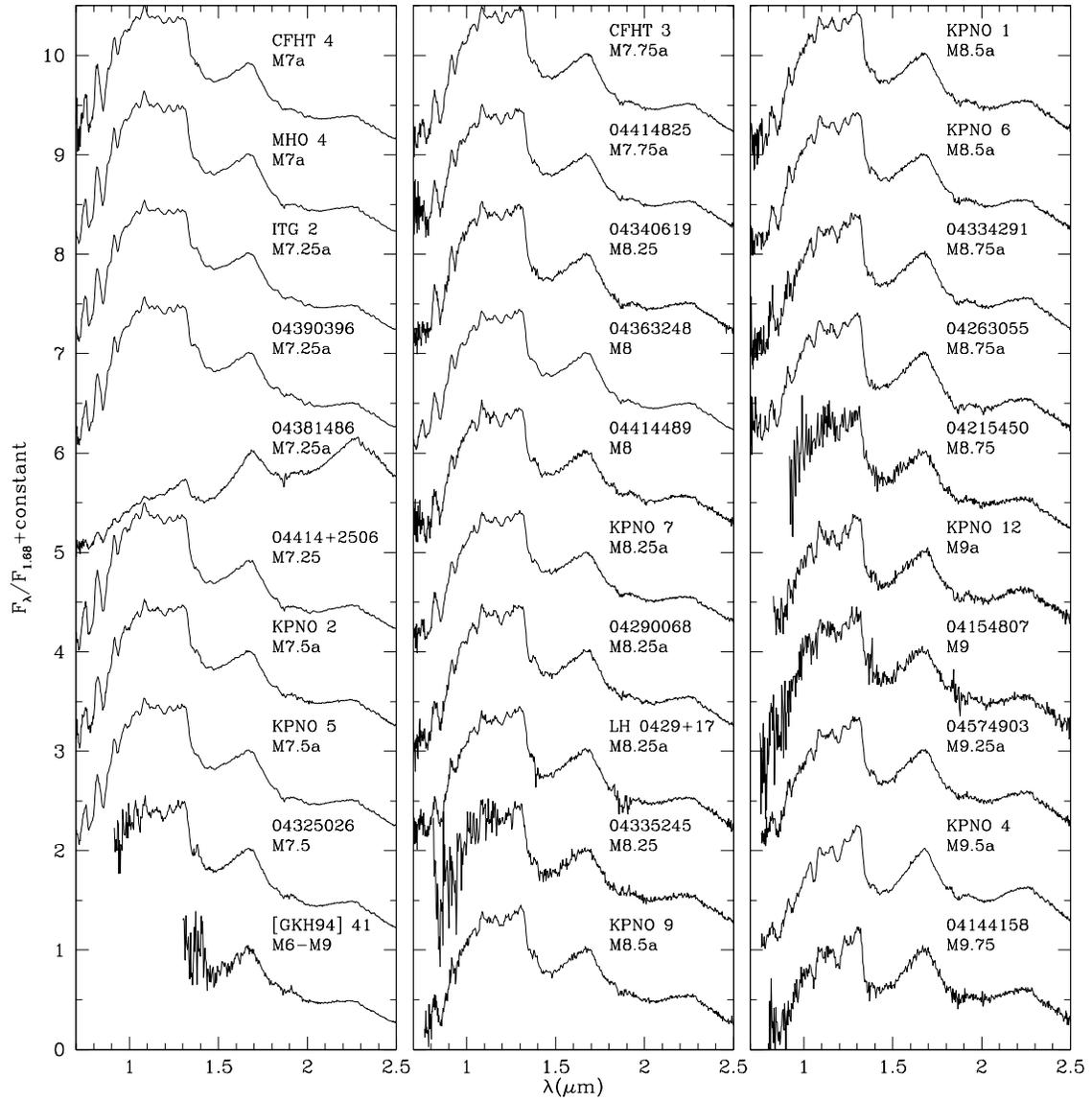}
\caption{
More near-IR spectra of members of Taurus (see Figure~\ref{fig:ir2}).
The spectrum of 2MASS~J04381486+2611399 has not been dereddened because
it is detected primarily in scattered light \citep{luh07edgeon}.
The data used to create this figure are available.
}
\label{fig:ir8}
\end{figure}

The optical spectra are presented in Figure~\ref{fig:op}, which apply to
one previously known member (FT Tau) and five new members.
The near-IR data for members are shown in
Figures~\ref{fig:ir1}--\ref{fig:ir8}, which consist of 19 spectra 
of new members and 180 spectra in 178 previously known systems
(both components of two binaries were observed). We have also included
the spectra of the two candidate companions to new members that have
uncertain membership (see Table~\ref{tab:mem}).
Four and 197 of those spectra were taken with NIRI and SpeX, respectively. 
The NIRI spectra are shown after smoothing to the resolution of the SpeX data.
We have included all SpeX data for Taurus members from our previous studies
\citep{luh06tau1,luh06tau2,luh07edgeon,luh09fu,luh09tau,mue07,esp14},
which correspond to 76 spectra in 75 systems.

\subsection{Spectral Classification}
\label{sec:class}

We have measured spectral types for the candidate members and previously
known members of Taurus in our spectroscopic sample and have assessed
the membership of the former. We have also estimated extinctions for
the members using the SpeX data, which are sensitive to extinction because
they span a large range of wavelengths (0.8--2.5~\micron).

The candidates can be divided into the following categories based
on their spectra and other properties:
highly redshifted emission lines (galaxies); absorption lines from
hydrogen and metals (early-type stars and giants behind Taurus);
red, featureless spectra and mid-IR excess emission (protostars in Taurus);
absorption bands from TiO, VO, and H$_2$O (M-type objects in the field
and in Taurus). To distinguish between field dwarfs and Taurus members
in the latter category, we used gravity sensitive absorption features
(Na~I, K~I, H$_2$O) and, when available, signatures of youth in the form
of strong emission lines and mid-IR excess emission.
As done in our previous studies \citep{luh99}, we measured spectral types
from the optical spectra of the young objects through comparison to field dwarfs
and averages of dwarfs and giants for $<$M5 and $\geq$M5, respectively
\citep{hen94,kir91,kir97}. The near-IR spectra were classified with standard
spectra from our previous work, \citet{cus05}, and \citet{ray09} for the field
dwarfs and with standard spectra from the Appendix for the young objects.
The errors for the optical and IR types are $\pm0.25$ and
0.5~subclass, respectively, unless indicated otherwise.
The spectral types for the new members and non-members are
listed in Tables~\ref{tab:new} and \ref{tab:non}, respectively.
The former are also included in the compilation of classifications for
all known members in Table~\ref{tab:mem}.
For previously known members that have SpeX classifications that are
more uncertain than those available from previous optical spectroscopy,
or that serve as near-IR classification standards (Appendix),
we do not report a spectral type from SpeX, and instead adopt the previous
optical type.

For Taurus members that have measured spectral types and that were
observed with SpeX, we have estimated extinctions by comparing the
observed spectral slopes at 1~\micron\ to the slopes of our young standards.
If the spectral type from SpeX was uncertain and a more accurate type
was available from optical data in a previous study, then we adopted that
classification when estimating the extinction. If the signal-to-noise (S/N)
at 1~\micron\ was low, then we estimated the extinction from the
slope at longer wavelengths if an absence of mid-IR excess emission indicated
that $K$-band excess was unlikely to be present.
The resulting extinctions were used for dereddening the spectra in
Figures~\ref{fig:ir2}--\ref{fig:ir8}.
For Taurus members that have measured spectral types but were not
observed by SpeX, we have estimated extinctions with the methods
described by \citet{fur11}, such as $J-H$ and $J-K$. Our extinction estimates
are presented in Table~\ref{tab:mem}. For some of the objects for which
we used the same methods as \citet{fur11}, our estimates differ
from their values because of differences in the adopted spectral types.
Our measurements of $A_J$ from the SpeX data have
uncertainties of $\sim$0.1~mag for objects that lack strong near-IR excess
emission from disks and that have low-to-moderate extinctions ($A_J\lesssim2$).
The uncertainties are larger for the remaining members, particularly
when strong excess emission is present throughout the near-IR bands.
Most of the SpeX extinctions agree with the values from the alternative
methods in \citet{fur11} to within $A_J<0.3$ and 0.1~mag for objects with
and without large near-IR excesses, respectively.

\subsection{Comments on Individual Sources}
\label{sec:comments}

{\it 2MASS J04345973+2807017}.
\citet{luh09tau} identified it as a candidate based on mid-IR excess emission.
They obtained a SpeX spectrum that was consistent with either a field
dwarf or a young star, but it appeared to agree better with the former.
However, the gravity sensitive features in our optical spectrum do
indicate that it is young. As noted in \citet{luh09tau}, it is much fainter
than other members near its spectral type, which indicates that it is
either a background young star that is not associated with Taurus or
a member that is seen in scattered light. We assume the latter for the
purposes of this work.

\begin{figure}
\epsscale{0.4}
\plotone{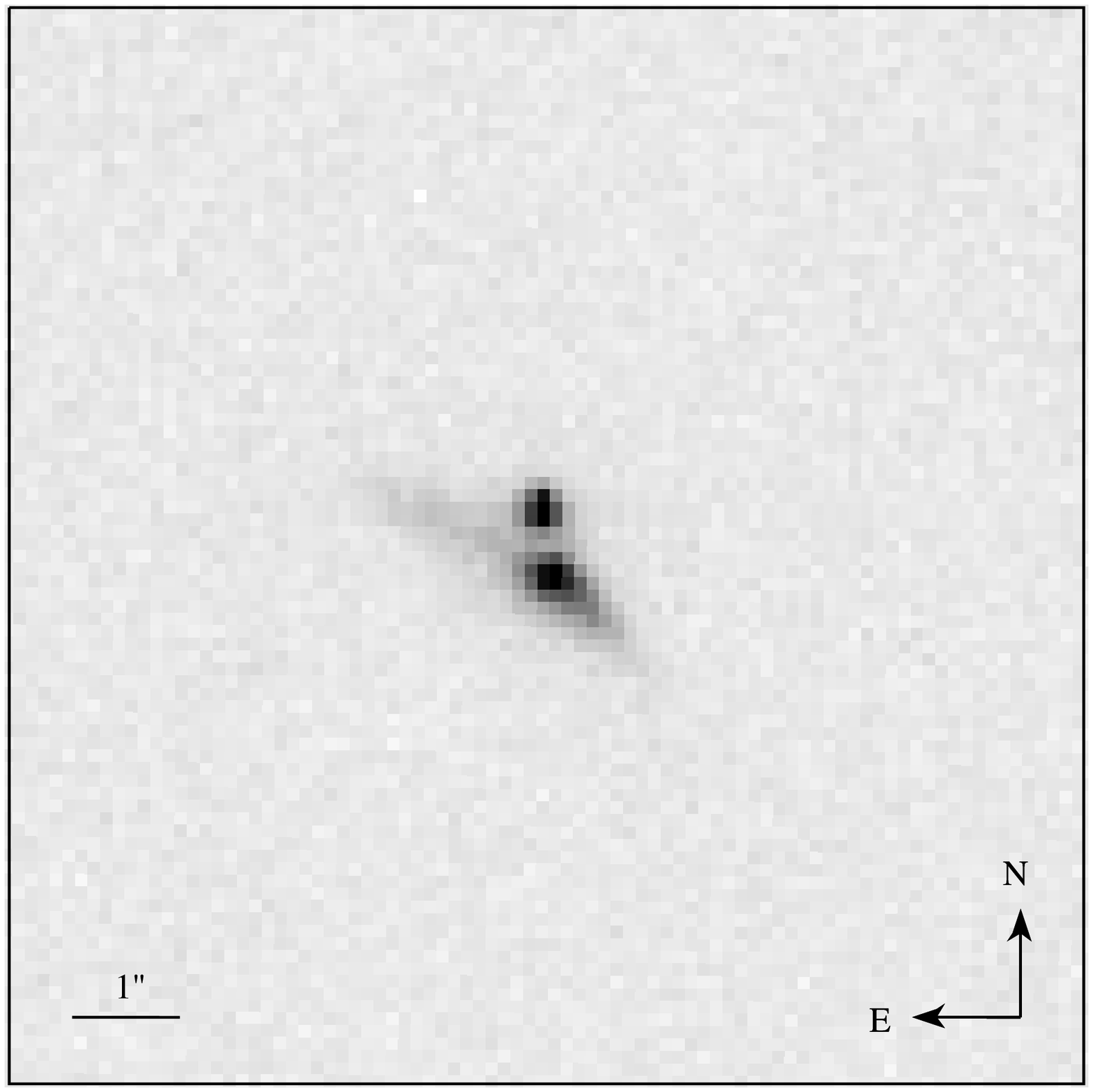}
\caption{
NIRI $K\arcmin$-band image of the components of 2MASS~04153452+2913469.
The size of the field is $10\arcsec\times10\arcsec$.
}
\label{fig:niri}
\end{figure}

{\it 2MASS J04153452+2913469}.
It was selected for spectroscopy based on its mid-IR excess in photometry
from {\it WISE} \citep{reb11,esp14}.
\citet{reb11} referred to it as a known galaxy, possibly because it is labeled
as such by SDSS. Our NIRI acquisition image obtained prior to spectroscopy
resolved it as a pair of sources that are separated by $0\farcs6$, as shown 
in Figure~\ref{fig:niri}. It is unclear which component dominates the mid-IR
flux detected by {\it WISE}. 
The spectrum of the northern component contains Br$\gamma$ emission and
CO absorption, which are indicative of a young low-mass star
(Figure~\ref{fig:ir1}).
The southern component exhibits extended emission that could be consistent
with either a galaxy or reflection nebulosity surrounding a young star.
If it is the latter, then it is probably a protostar based on its
red, featureless spectrum.

\section{Initial Mass Function}
\label{sec:imf}

\subsection{Completeness in the SDSS Field}

\begin{figure}
\epsscale{1.1}
\plotone{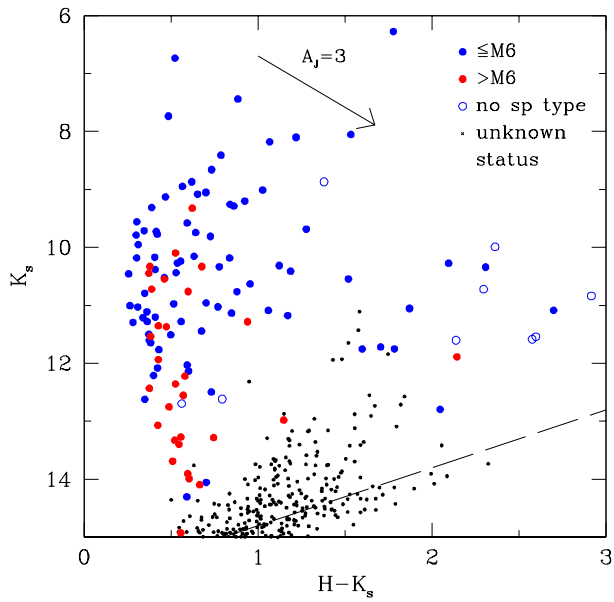}
\caption{
CMD from 2MASS for the known members of Taurus
(large filled circles) and other sources with unconstrained membership
(small points) that are within the SDSS field from \citet{fin04}
(Figure~\ref{fig:map}). We have omitted stars that are likely non-members
based on CMDs (Figure~\ref{fig:cmd}), proper motions, or spectroscopy.
The completeness limit of the 2MASS data is indicated (dashed line).
}
\label{fig:hk}
\end{figure}

We have attempted to perform a thorough survey for new members of Taurus
within the SDSS field from \citet{fin04} in an effort to measure the IMF
across a large portion of this region.
To construct an IMF from our updated census of Taurus members within the
SDSS field, we have begun by characterizing the completeness of that census.
As in our previous IMF studies of this kind, we have investigated the
completeness using a near-IR CMD because it is sensitive to members at low
masses and high extinctions.
In Figure~\ref{fig:hk}, we show a diagram of $K_s$ versus $H-K_s$ based on
2MASS data for the known members of Taurus and other sources with
unconstrained membership within the SDSS field. We have omitted objects that
are likely to be non-members based on the CMDs in Figure~\ref{fig:cmd},
proper motions, or spectroscopy from this work and previous studies.
The CMD in Figure~\ref{fig:hk} indicates that the census of Taurus members
within the SDSS field from \citet{fin04} has a high degree of completeness
for large ranges of magnitude and extinction. 
Therefore, a sample of known members in the SDSS field that have
extinctions below a given threshold should be representative of the
IMF in that field down to a certain mass, assuming that the average
extinction does not vary with mass. For our IMF sample, we have selected
a threshold that is high enough to encompass a large number of members while
low enough that the sample has a high level of completeness down substellar
masses, arriving at $A_J<3$. Based on Figure~\ref{fig:hk}, the current
census in the SDSS field is nearly complete for extinction-corrected
magnitudes of $K_s<12.7$ within that threshold, which corresponds to masses
of $\lesssim0.03$~$M_\odot$ for ages of a few Myr according to evolutionary
models \citep{bar98,bar15}.

\subsection{Comparison to XEST Fields and IC 348}

The completeness limit of 0.03~$M_\odot$ for our extinction-limited
sample in the SDSS field is similar to the limit of 0.02~$M_\odot$ for the IMF
in the XEST fields from \citet{luh09tau}. As a result, the two mass functions
can be directly compared. In particular, we would like to check whether the
surplus of solar-mass stars in the XEST fields relative to other
star-forming regions is also present across a larger area like
the SDSS field (see Section~\ref{sec:intro}).

Masses of young stars are typically derived by combining estimates
of bolometric luminosities (from photometry) and effective temperatures
(from spectral types) with the values predicted by evolutionary models.
However, because these mass estimates are subject to uncertainties in
the adopted bolometric corrections, temperature scales, and models,
and because we are primarily interested in detecting variations in the
IMF between different samples of young stars, we have chosen to
use the distributions of spectral types in those samples as observational
proxies for their IMFs.
The ages of Taurus and IC~348 derived from comparisons of their
low-mass stars to theoretical isochrones are sufficiently similar 
\citep[$\sim$1 and 2~Myr,][]{luh03ic} that
a given spectral type should correspond to the same mass in both regions.
It has been suggested that isochronal ages for
star-forming regions may be underestimated by a factor of $\sim2$
\citep{nay09,bel13}. Even if that is true for one of the two regions,
the relationships between spectral types and masses should still be nearly
identical between the regions for masses of $\lesssim1.5$~$M_\odot$ according
to evolutionary models \citep{bar98}.

\begin{figure}
\epsscale{1}
\plotone{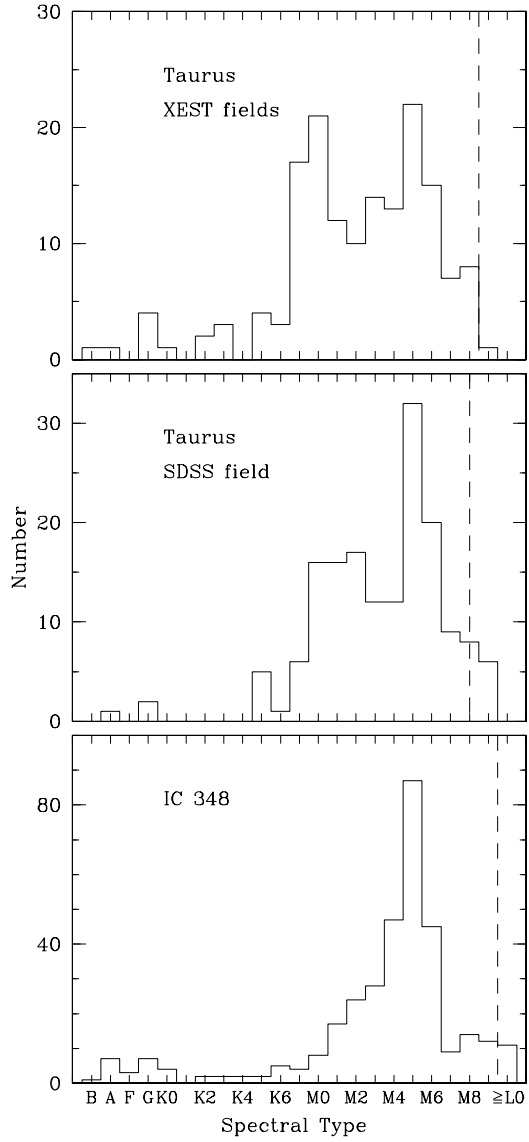}
\caption{
Distributions of spectral types for known members of Taurus within
the XEST fields \citep{luh09tau}, Taurus members with $A_J<3$ in the SDSS
field from \citet{fin04}, and members of IC~348 with $A_J<1.5$ in
a field encompassing most of that cluster \citep{luh16}.
The completeness limits of these samples of members are indicated
(dashed lines).
}
\label{fig:histo}
\end{figure}

In Figure~\ref{fig:histo}, we show the spectral type distributions for
our extinction-limited sample in the SDSS field. For comparison,
we also include the IMF sample in the XEST fields \citep{luh09tau} and
an extinction-limited sample in IC~348 \citep{luh16}, which is representative
of IMFs in nearby star-forming clusters \citep{hil97,luh07cha}.
The sample of members in the SDSS field exhibits a surplus of 
stars near M0 ($\sim0.7$~$M_\odot$) relative to IC~348, but it is
smaller than that in the XEST fields.
This indicates that the solar-mass stars are slightly concentrated near
the aggregates, and that the IMF on large scales in Taurus does differ
from denser clusters. Thus, it appears that previously reported differences
in IMFs between Taurus and denser clusters were a reflection of modest
levels of both mass segregation and IMF variations.

\subsection{Effects of Magnetic Activity}

We now discuss a recently proposed source of systematic error in
measurements of IMFs in Taurus and other star-forming regions.
The young eclipsing binary brown dwarf 2MASS J05352184$-$0546085 exhibits
an anomaly in which the primary is cooler than the secondary \citep{sta06},
which has been attributed to a reduction of the primary's temperature by
magnetic activity \citep{rei07,sta07}.
If the spectral types of young stars and brown dwarfs can be significantly
affected by activity, then their mass estimates, and hence their IMFs,
may contain large systematic errors \citep{moh09}.
\citet{sta12,sta14} attempted to quantify those errors, concluding that the
masses derived from spectral types could be underestimated by up to a
factor of two. Such error estimates depend on the adopted conversion between
spectral types and temperatures; a temperature scale for inactive stars
was used in those studies. However, if one instead derives masses with
a scale for active stars, the systematic errors would be much smaller.
For instance, the widely adopted scale from \citet{luh03ic} was designed for use
with young stars and brown dwarfs, and thus may account for the typical
effects of activity in such populations.
Systematic errors can arise at multiple stages in the derivation of masses
for young stars (e.g., temperature scale, evolutionary models), regardless
of whether activity is responsible for some of those errors. 
The most important question is whether the combination of those various 
stages produce masses that are accurate \citep{luh12}.
As discussed by \citet{luh06abdor} and \citet{luh12},
the IMFs of low-mass stars and brown dwarfs that we have previously derived
for star-forming clusters are consistent with mass functions of open clusters
\citep{mor04} and the solar neighborhood \citep{rei02,boc10,kir12},
and thus show no evidence of large systematic errors.

\section{Conclusions}

We have attempted to measure the IMF across a larger fraction of Taurus
than considered in previous IMF studies of this region to better determine
whether it exhibits an anomalous IMF relative to denser nearby clusters
like IC~348. To do this, we have performed a thorough survey for new members
within a $\sim$40~deg$^2$ field that was imaged by SDSS. We have obtained
spectra of candidate members appearing in CMDs for that field,
as well as a miscellaneous sample of candidates across all of Taurus that
were selected with a variety of diagnostics of membership.
Through our spectroscopy, we have classified 22 candidates as new members,
which includes one of the coolest known members (M9.75).
The update census of Taurus now contains 413 members that are resolved
by 2MASS or {\it Spitzer}. For the SDSS field, we have constructed an
extinction-limited sample of members that should be nearly complete down to
masses of $\sim0.03$~$M_\odot$. That sample exhibits a surplus of solar-mass
stars relative to clusters like IC~348, although it is less pronounced than in
previously reported IMFs in Taurus that have been measured for smaller fields. 
{\it Gaia} \citep{per01} and Pan-STARRS1 \citep{kai02} will soon provide
photometry and astrometry that can be used to search the entire extent
of the Taurus cloud complex for new members down to substellar masses and
at low-to-moderate extinctions, which will further improve the statistical
accuracy of the measurement of the IMF.

\acknowledgements
K. L. acknowledges support from NSF grant AST-1208239 and
E. M. acknowledges support from NSF grant AST-1313029 and the NASA NExSS
program. We thank Jackie Faherty for providing her SpeX data and
David Schlegel and Doug Finkbeiner for information regarding the SDSS data.
The Gemini data were obtained through program GN-2009B-Q-91.
Gemini Observatory is operated by AURA under a cooperative agreement with
the NSF on behalf of the Gemini partnership: the NSF (United States), the NRC
(Canada), CONICYT (Chile), the ARC (Australia),
Minist\'{e}rio da Ci\^{e}ncia, Tecnologia e Inova\c{c}\~{a}o (Brazil) and
Ministerio de Ciencia, Tecnolog\'{i}a e Innovaci\'{o}n Productiva (Argentina).
The IRTF is operated by the University of Hawaii under contract 
NNH14CK55B with NASA. 
2MASS is a joint project of the University of Massachusetts and IPAC
at Caltech, funded by NASA and the NSF. 
Funding for SDSS has been provided by the Alfred P. Sloan Foundation,
the Participating Institutions, the NSF,
the U.S. Department of Energy, NASA, the Japanese Monbukagakusho, the
Max Planck Society, and the Higher Education Funding Council for England.
The SDSS Web Site is http://www.sdss.org/.
The SDSS is managed by the Astrophysical Research Consortium for the
Participating Institutions. The Participating Institutions are the American
Museum of Natural History, Astrophysical Institute Potsdam, University of
Basel, University of Cambridge, Case Western Reserve University, The University
of Chicago, Drexel University, Fermilab, the Institute for Advanced Study, the
Japan Participation Group, The Johns Hopkins University, the Joint Institute
for Nuclear Astrophysics, the Kavli Institute for Particle Astrophysics and
Cosmology, the Korean Scientist Group, the Chinese Academy of Sciences,
Los Alamos National Laboratory, the Max-Planck-Institute for Astronomy,
the Max-Planck-Institute for Astrophysics, New Mexico State University,
Ohio State University, University of Pittsburgh, University of Portsmouth,
Princeton University, the United States Naval Observatory, and the University
of Washington. The HET is a joint project of
the University of Texas at Austin, the Pennsylvania State University,
Stanford University, Ludwig-Maximillians-Universit\"at M\"unchen, and
Georg-August-Universit\"at G\"ottingen. The HET is named in honor of
its principal benefactors, William P. Hobby and Robert E. Eberly. The
Marcario Low-Resolution Spectrograph at HET is named for Mike Marcario
of High Lonesome Optics, who fabricated several optics for the
instrument but died before its completion; it is a joint project of
the HET partnership and the Instituto de Astronom\'{\i}a de la
Universidad Nacional Aut\'onoma de M\'exico. 
WIRCam is a joint project of CFHT, Taiwan, Korea, Canada, and France.
CFHT is operated by the NRC of Canada, the Institute
National des Sciences de l'Univers of the Centre National de la Recherche
Scientifique of France, and the University of Hawaii.
This work used data from the SpeX Prism Spectral Libraries (maintained
by Adam Burgasser at http://www.browndwarfs.org/spexprism), the NASA/IPAC
Infrared Science Archive (operated by JPL under contract with NASA),
and SIMBAD database, operated at CDS, Strasbourg, France.
The Center for Exoplanets and Habitable Worlds is supported by the
Pennsylvania State University, the Eberly College of Science, and the
Pennsylvania Space Grant Consortium.

\appendix

\section{Near-IR Spectral Standards for M and L Types at Young Ages}

\subsection{Selection Criteria}

If field dwarfs are used as the standards when measuring spectral types of
young late-type objects at near-IR wavelengths, the resulting types
are often systematically later than those that would be derived with optical
spectroscopy because the depths of the strongest near-IR features, the steam
bands, depend on surface gravity \citep{luh03ic,luh12}.
In our previous studies, to measure IR spectral types 
that are tied to those at optical wavelengths,
we have classified the IR spectra of young M-type stars
via comparison to other young objects for which we have measured optical types.
The number of young stars with both optical spectral types and IR spectra
is now large enough that we can combine the spectra of several objects
at each subclass for use as classification standards.  To do this for ages
of $\lesssim10$~Myr, we consider objects that satisfy the following criteria:
classified at optical wavelengths with our methods for M types or with
those from \citet{cru07,cru09,cru16} and \citet{kir10} for L types;
member of a nearby cluster or association that has an age of $\lesssim10$~Myr
or a member of the solar neighborhood that has an age of $\lesssim100$~Myr
based on gravity-sensitive spectral features;
unlikely to have $K$-band excess emission from circumstellar dust based
on near- to mid-IR photometry; and SpeX spectra that have S/N$\gtrsim30$
and good telluric correction are available.
Because of their uncertain ages, objects in the solar neighborhood
are used only for spectral types in which there are few members of clusters
that have accurate optical spectral types and adequate spectra, which
corresponds to the L types. The surface gravities indicated by optical
spectra have been previously denoted by suffixes of $\alpha$ (highest gravity),
$\beta$, $\gamma$, and $\delta$ (lowest gravity) that are appended to spectral
types \citep{kir05,kir06,cru09}. We have considered L dwarfs classified
as $\gamma$ or $\delta$, which likely correspond to ages of $\sim10$--100
and $\lesssim10$~Myr, respectively.

\subsection{New Data in TWA}

Members of the TW Hya association (TWA) are especially appealing 
as spectral standards since they are nearby and are just old enough to
no longer be obscured by their natal molecular cloud \citep[30--80~pc,
$\sim10$~Myr,][]{web99,mam05}. We have observed many of those objects
with optical spectroscopy to obtain optical spectral types that have been
measured with our methods, facilitating their inclusion in our new standards.
These data were taken with the Goodman High Throughput Spectrograph at the
Southern Astrophysical Research (SOAR) Telescope on the nights of 2014 May 8
and 9 (TWA 1, 2, 3A, 3B, 4, 5A, 6, 7, 8A, 8B, 10, 11C, 12, 13, 16, 20, 23, 25,
30, 32) and the Cerro Tololo Ohio State Multi-Object Spectrograph
(COSMOS) at the 4~m Blanco telescope at Cerro Tololo Inter-American
Observatory on the night of 2015 May 12 (TWA 26, 33, 34).
Goodman was operated with the 400~l/mm grating in second order,
the GG495 filter, and the $0\farcs46$ slit (3~\AA\ resolution).
COSMOS was configured with the red VPH grism, the OG530 filter, and slit
widths of $0\farcs9$ (TWA 33 and 34, 3~\AA) and $1\farcs2$ (TWA 26, 4~\AA). 
The wavelength coverage was roughly 5500--9500~\AA\ for both
instruments. In Table~\ref{tab:twa}, we have compiled previous spectral types
and our new measurements for the known members of TWA. For TWA 29, we
have included the spectral type that we have measured from the optical spectrum
collected by \citet{loo07}, which is nearly the same as one reported in that
study. To provide the near-IR spectra needed for our standards, we also
observed some of the members of TWA with the prism mode of SpeX ($0\farcs8$
slit) on the nights of 2005 December 14 (TWA 1, 2, 5, 6, 7, 8A, 8B, 26, 27, 28),
2011 December 3 (TWA 13), and 2015 April 19--22 (TWA 10, 12, 11C, 23, 32,
33, 34).

\subsection{Construction of Standard Spectra}
\label{sec:construction}

\begin{figure}
\epsscale{0.55}
\plotone{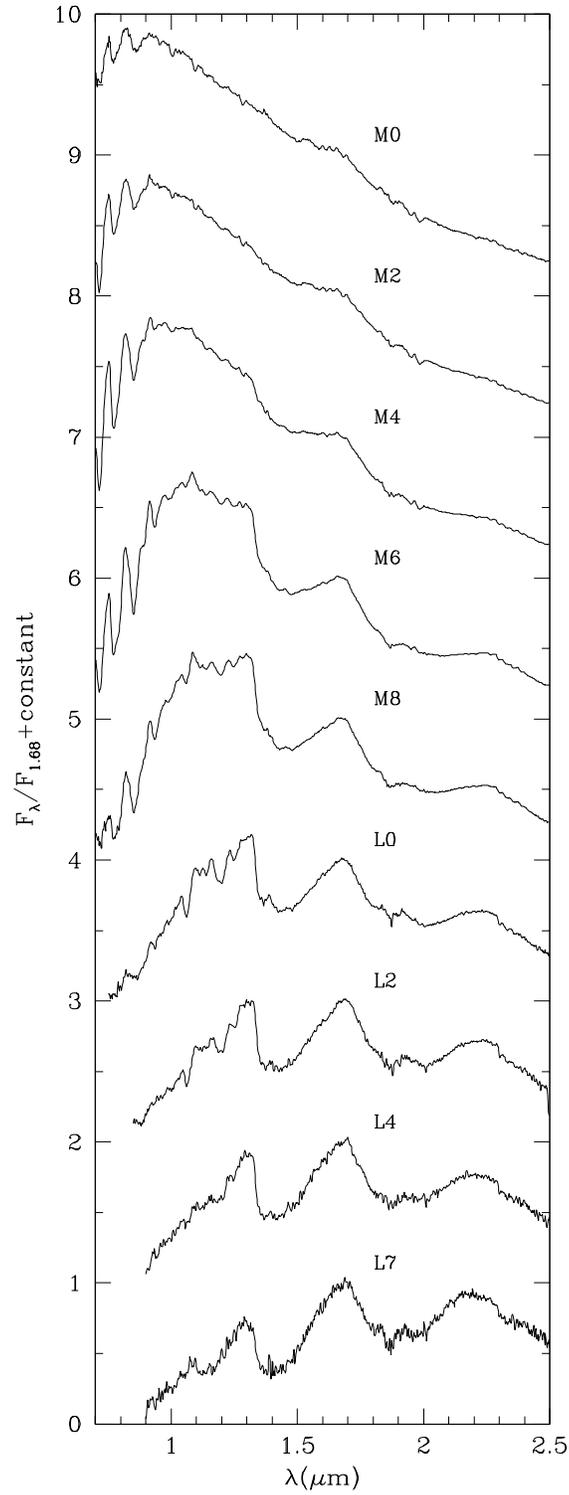}
\caption{
Standard spectra for classifying young stars and brown dwarfs, which
have been constructed from the SpeX data listed in Table~\ref{tab:stand}.
The data used to create this figure are available.
}
\label{fig:stand1}
\end{figure}

\begin{figure}
\epsscale{1}
\plotone{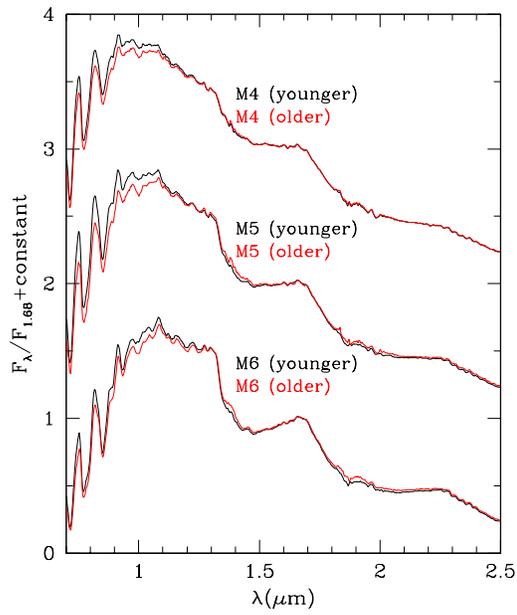}
\caption{
Comparisons of the standard spectra constructed from members
of younger ($\lesssim5$~Myr) and older ($\sim10$~Myr) clusters
and associations. On average, the two populations exhibit different
0.7--1.1~\micron\ slopes at mid-M spectral types.
}
\label{fig:stand2}
\end{figure}

After compiling SpeX data from this work and previous studies for objects
that satisfied our selection criteria, we adjusted the slope of each spectrum
among the M types using the reddening law of \citet{car89} so that its $J-H$
color agreed with the typical intrinsic value for a young star at its optical
spectral type \citep{luh10tau}. This was done to correct for reddening by
interstellar dust and to facilitate comparison of all spectra near a given
spectral type.
The spectral slopes of the L dwarfs were not dereddened since they reside in
the solar neighborhood, and thus should have little extinction, and since the
intrinsic colors of L dwarfs are not well-determined for the age range in
which we are attempting to create spectral standards ($\lesssim10$~Myr). 
We then compared the spectra of objects with similar optical types and
rejected spectra that were outliers in the strengths of their absorption
features relative to most other data. The unrejected spectra within
bins of 0.5--1~subclass were combined with the appropriate weights based
on their optical spectral types to produce a spectrum at each 0.5~subclass
interval between M0--L0 and at L2, L4, and L7. We compared the spectra
between younger and older populations (Taurus and IC~348 versus TWA and
Upper Sco) and found that the 0.7--1.1 slopes were systematically redder
at older ages for types of M3.5--M6.5.
As a result, the spectra of the two populations were split into two 
standard spectra at each spectral type within that range.
In Table~\ref{tab:stand}, we list the objects used to create the standard
spectra, the optical spectral types that were adopted for them, and dates
of their SpeX observations.
We show a sample of the standard spectra in Figure~\ref{fig:stand1}.
The displayed spectra for M4 and M6 apply to the younger populations.
Electronic versions of all of the standard spectra are available via 
Figure~\ref{fig:stand1}. The differences in the 0.7--1.1 slopes between the
younger and older objects are illustrated in Figure~\ref{fig:stand2}, where 
we compare the standard spectra for the two populations at M4, M5, and M6.

Because the young L dwarfs that we considered exhibit large variations in
their spectral slopes and poorly populate some subclasses, we describe
for the individual L subclasses the spectra that we chose to use as standards
and those that we rejected. In this discussion, we quantify differences
in the slopes in the SpeX data between pairs of objects in terms of the amount
of extinction, $A_V$, that would produce the same difference via reddening.
At L0, 2MASS J01415823$-$4633574, 2MASS J22134491$-$2136079,
2MASS J23153135+0617146 have similar features and slopes in their SpeX data
while 2MASS J02411151$-$0326587 is bluer by $A_V=0.5$ and
2MASS J03231002$-$4631237 is redder by $A_V=0.7$. Because of the agreement of
the first three spectra, we adopted their mean spectrum as the standard for L0.
For the one known member of Taurus with an optical spectral type of L0,
2MASS J04373705+2331080 \citep{luh09tau}, the SpeX spectrum from \citet{bow14}
agrees with our L0 standard except that the Taurus object is redder by $A_V=2$,
which is likely due to some combination of extinction and an intrinsically
redder color. The latter is plausible given that it is likely younger
than the field L dwarfs we have considered, and L dwarfs tend to have redder
near-IR colors at younger ages \citep{kir06,cru09}.
At L1, only 2MASS J05184616$-$2756457 satisfied our criteria for
consideration as a standard, and its spectrum differs only slightly from
the L0 standard in the depths of its steam bands and its spectral slope.
Therefore, we have not adopted a standard spectrum for that spectral type.
At L2, 2MASS 05361998-1920396 and 2MASS 00550564+0134365 have
similar spectra except for a modest difference in their spectral slopes
($A_V=0.4$), so we have adopted the mean of their spectra as the L2 standard.
Among the L4 dwarfs that we examined, 2MASS J15382417$-$1953116
has the bluest spectrum, and 2MASS J05012406$-$0010452, 2MASS J15515237+0941148,
2MASS J01262109+1428057, and 2MASS J16154255+4953211 are redder by
$A_V=0.4$, 0.4, 1.8, and 3, respectively. The steam bands are slightly weaker
in the latter two objects than in the others. We have adopted the mean
of 2MASS J05012406$-$0010452 and 2MASS J01262109+1428057 as the L4 standard.
The one L3$\gamma$ dwarf with a SpeX spectrum, 2MASS J22081363+2921215, 
is bluer than our adopted L2 standard, and thus does not follow the trend
of redder spectra with later L types, making it unsuitable as a standard.
The only two L5 dwarfs that we examined are 2MASS J05120636$-$2949540 
and 2MASS J03552337+1133437; the former is similar to the L4 dwarfs and
the latter is an outlier in its slope and the depths of its steam bands,
so we have not adopted a standard at L5.
Since optical spectral types have not been defined for young
L dwarfs later than L5 \citep{cru09}, we have adopted the near-IR types of L7
that have been proposed for TWA 41 and 42 \citep{kel15,sch16} and we take
the mean of their spectra as the standard at that type. We note that
the near-IR spectrum of the young late-L dwarf PSO J318.5338$-$22.8603
\citep{liu13} is similar to the data for those TWA members except that
it is redder by $A_V=1.5$.

\subsection{Application of the Standard Spectra}

\begin{figure}
\epsscale{1}
\plotone{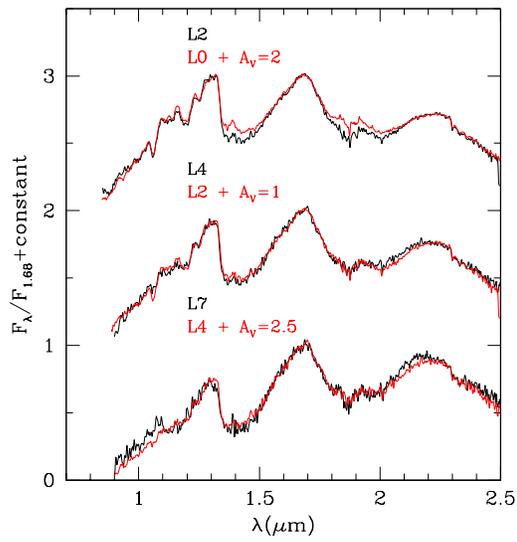}
\caption{
Comparisons of standard spectra at L2, L4, and L7 to earlier standards
that have been reddened to the same spectral slope. The spectra within
each pair are quite similar, which indicates that
IR spectral types measured for reddened young L dwarfs (i.e., members of
star-forming regions) can have large uncertainties.
}
\label{fig:stand4}
\end{figure}

Our standard spectra have been designed for classifying M- and L-type objects
that have ages of $\lesssim10$~Myr. Most objects at these ages reside
in clusters and associations that are embedded within molecular clouds.
Populations like Upper Sco that are no longer associated with
molecular clouds also can exhibit noticeable reddening ($A_V>0.2$) from the
interstellar medium beyond the Local Bubble ($d\gtrsim100$~pc).
As a result, to classify an object's spectrum with our standard spectrum,
one should identify the combination of spectral type and reddening that
provides the best fit. Even in the presence of reddening, it is normally
possible to measure IR types with errors of only $\pm0.5$ at M types.
However, because the depths of the steam bands do not vary significantly among
the L types, there is greater degeneracy between spectral type and 
reddening, resulting in larger uncertainties in the two parameters.
The degeneracy is illustrated in Figure~\ref{fig:stand4}, which shows
that the unreddened standards at L2, L4, and L7 are similar to reddened
versions of the standards at L0, L2, and L4, respectively.
Because young L dwarfs at a given optical spectral type exhibit a fairly
large range of spectral slopes (Section~\ref{sec:construction}), a similar
degeneracy is present in the IR classifications of young L dwarfs in the
solar neighborhood.

\begin{figure}
\epsscale{1}
\plotone{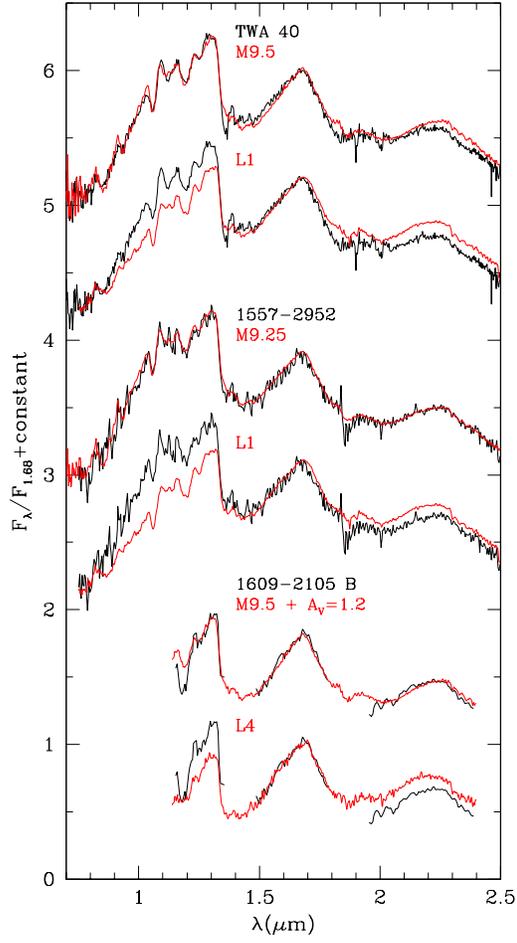}
\caption{
Comparisons of spectra of three young late-type objects
\citep{kir10,laf08,laf10,gag14} to our best-fitting standard spectra (M9.5,
M9.25, M9.5) and our standards for the spectral types from previous studies
(L1, L1, L4).
}
\label{fig:stand3}
\end{figure}

We have applied our standard spectra to a sampling of IR spectra that have
been classified in other studies. We frequently arrive at types that
are earlier than the previous results.
Examples of three such objects
are shown in Figure~\ref{fig:stand3}, where we compare their spectra to our
best-fitting standards and our standards for the previously reported types.
The first object, TWA~40, was previously classified as L0 and L1 at optical
and IR wavelengths, respectively \citep{gag14,gag15}. We find that
our M9.5 standard provides the best match to its IR spectrum.
The IR type from \citet{gag15} was partially based on a comparison
to members of Upper Sco that had been classified as L0--L1 with IR data
\citep{lod08}. 
However, optical spectroscopy indicates that several of those sources in
Upper Sco have spectral types near M9 \citep[][K. Luhman,
in preparation]{her09}, which would explain why \citet{gag15} found a type
later than our value.
The second object in Figure~\ref{fig:stand3}, 2MASS J15575011$-$2952431,
is a similar example of a source previously classified as early L
\citep{all13,fah16} for which we measure a type near M9.
Our IR classification agrees with the optical spectral type of that object
\citep{kir10}. Finally, we compare in Figure~\ref{fig:stand3} the IR spectrum
of 1RXS J160929.1$-$210524~B \citep{laf08,laf10} to our best-fitting standard
(M9.5 with $A_V=1.2$) and our standard for the type of L4 that was measured by
\citet{laf08} and \citet{lac15}. The companion agrees well with M9.5, and is
too blue to be a typical young L4 dwarf. The extinction estimate
from our classification is lower than the value of $A_V=4.5$ derived
by \citet{wu15} through a comparison of the observed near-IR spectral energy
distribution to model predictions.

\clearpage

\begin{deluxetable}{ll}
\tabletypesize{\scriptsize}
\tablewidth{0pt}
\tablecaption{Members of Taurus\label{tab:mem}}
\tablehead{
\colhead{Column Label} &
\colhead{Description}}
\startdata
2MASS & 2MASS Point Source Catalog source name \\
WISE & WISE All Sky Source Catalog source name\tablenotemark{a} \\
RAh & Hour of Right Ascension (J2000) \\
RAm & Minute of Right Ascension (J2000) \\
RAs & Second of Right Ascension (J2000) \\
DEd & Degree of Declination (J2000) \\
DEm & Arcminute of Declination (J2000) \\
DEs & Arcsecond of Declination (J2000) \\
Name & Source name \\
SpType & Spectral type \\
r\_SpType & Spectral type reference\tablenotemark{b} \\
Adopt & Adopted spectral type \\
Spexdate & SpeX date \\
Aj & Extinction in $J$ \\
f\_Aj & Method of extinction estimation\tablenotemark{c}
\enddata
\tablenotetext{a}{Source names for HBC 360, HBC 361, and IRAM 04191+1522 are from the ALLWISE Source Catalog.}
\tablenotetext{b}{
(1) \citet{hk03};
(2) \citet{her14};
(3) \citet{luh09tau};
(4) \citet{kh95};
(5) \citet{whi04};
(6) \citet{dop05};
(7) \citet{pra09};
(8) \citet{esp14};
(9) \citet{wic96};
(10) \citet{ngu12};
(11) \citet{luh06tau1};
(12) this work;
(13) \citet{fin10};
(14) \citet{con10};
(15) \citet{tor95};
(16) \citet{sch09};
(17) \citet{reb10};
(18) \citet{luh04tau};
(19) \citet{wel95};
(20) \citet{whi03};
(21) \citet{bri98};
(22) \citet{str94};
(23) \citet{sce08};
(24) \citet{bri02};
(25) \citet{gui06};
(26) \citet{sle06};
(27) \citet{lr98};
(28) \citet{luh03tau};
(29) \citet{luh98vx};
(30) \citet{mar01};
(31) \citet{har94};
(32) \citet{ken98};
(33) \citet{bec07};
(34) \citet{luh99};
(35) \citet{cal04};
(36) \citet{rei99};
(37) \citet{luh06tau2};
(38) \citet{cie12};
(39) \citet{duc99};
(40) \citet{whi01};
(41) \citet{bon14};
(42) \citet{ito05};
(43) \citet{mon98};
(44) \citet{muz03};
(45) \citet{kra09};
(46) \citet{whi99};
(47) \citet{mar99};
(48) \citet{pra02};
(49) \citet{wal03};
(50) \citet{cow72};
(51) \citet{mar94};
(52) \citet{her08};
(53) \citet{nes95};
(54) \citet{mar00};
(55) \citet{bow15};
(56) \citet{bri99};
(57) \citet{har95};
(58) \citet{coh79};
(59) \citet{bos37};
(60) \citet{ste01};
(61) \citet{mal98};
(62) \citet{bri93}.
}
\tablenotetext{c}{
$J-H$ and $J-K$ = derived from these colors assuming photospheric
near-IR colors; CTTS = derived from $J-H$ and $H-K$ assuming
intrinsic colors of classical T Tauri stars from \citet{mey97}; opt spec =
derived from an optical spectrum; SpeX = derived from a SpeX spectrum;
(1) = \citet{bri98};
(2) = \citet{luh00};
(3) = \citet{str94};
(4) = \citet{bec07};
(5) = \citet{whi01};
(6) = \citet{dew03};
(7) = \citet{cal04}.}
\tablecomments{
The table is available in a machine-readable form.}
\end{deluxetable}

\begin{deluxetable}{lllllll}
\tabletypesize{\scriptsize}
\tablewidth{0pt}
\tablecaption{New Members of Taurus in Spectroscopic Sample\label{tab:new}}
\tablehead{
\colhead{} &
\colhead{Spectral} &
\colhead{Basis of} &
\colhead{Telescope \&} &
\colhead{} \\
\colhead{2MASS} &
\colhead{Type} &
\colhead{Selection\tablenotemark{a}} &
\colhead{Instrument} &
\colhead{Date}}
\startdata
   J04105425+2501266 &   M1--M3 &      IR & Gemini/NIRI & 2009 Dec 26 \\
   J04144158+2809583 &   M9.75 &  WIRCam/IRAC &  IRTF/SpeX & 2013 Jan 3 \\
   J04153452+2913469 &   K-M+? &      IR & Gemini/NIRI & 2009 Dec 7 \\
   J04153566+2847417 & M4.5-M6.5 &      IR &  IRTF/SpeX & 2010 Jan 3 \\
   J04154131+2915078 &   M5.75 &    CMDs &  IRTF/SpeX & 2010 Jan 3 \\
   J04154269+2909558\tablenotemark{b} &    M6.5 &  companion? &    HET/LRS & 2009 Nov 8 \\
   J04154807+2911331 &      M9 &    CMDs &  IRTF/SpeX & 2010 Jan 3 \\
   J04161726+2817128 &   M4.75 &    CMDs &  IRTF/SpeX & 2010 Jan 16 \\
   J04213965+2649143 &      M6 &    CMDs &  IRTF/SpeX & 2010 Jan 3 \\
   J04215482+2642372 &      M5 &    CMDs &  IRTF/SpeX & 2010 Jan 4 \\
   J04245021+2641006 &      M5 &    CMDs &  IRTF/SpeX & 2010 Jan 4 \\
   J04264449+2756433 &      M6 &    CMDs &  IRTF/SpeX & 2010 Jan 4 \\
   J04272467+2624199 &      M3 &    CMDs &  IRTF/SpeX & 2010 Jan 16 \\
   J04314644+2506236 &    M5.5 &    CMDs &  IRTF/SpeX & 2010 Jan 4 \\
   J04340619+2418508 &   M8.25 &    CMDs &  IRTF/SpeX & 2010 Jan 3 \\
   J04344586+2445145 &   M4.75 &    CMDs &  IRTF/SpeX & 2010 Dec 23 \\
   J04345973+2807017 &   M5.75 &      IR &    HET/LRS & 2009 Nov 8 \\
   J04351316+1725496 & M4.5,M4 &      pm & HET/LRS,IRTF/SpeX & 2009 Dec 26,2010 Jan 3 \\
   J04355881+2438404 &      M3 &    CMDs &  IRTF/SpeX & 2010 Jan 16 \\
   J04363248+2421395 &      M8 &    CMDs &  IRTF/SpeX & 2010 Jan 4 \\
   J04590305+3003004 &      M2 & H$\alpha$,X-ray,pm &    HET/LRS & 2009 Dec 26 \\
  J05000310+3001074\tablenotemark{c} &    M1.5+? &   H$\alpha$,IR,pm & HET/LRS,IRTF/SpeX & 2009 Dec 26,2010 Jan 3 
\enddata
\tablenotetext{a}{Sources were selected for spectroscopy because
they were candidate members based on possible companionship to known
Taurus members, CMDs, X-ray emission, H$\alpha$ emission,
mid-IR excess emission, proper motions (``pm"), or red WIRCam/IRAC colors.}
\tablenotetext{b}{Candidate companion to IRAS~04125+2902.}
\tablenotetext{c}{GZ Aur A and B. Both components were observed by SpeX.
Only the primary was observed by LRS.}
\end{deluxetable}

\begin{deluxetable}{lllllll}
\tabletypesize{\scriptsize}
\tablewidth{0pt}
\tablecaption{Nonmembers in Spectroscopic Sample\label{tab:non}}
\tablehead{
\colhead{} &
\colhead{Spectral} &
\colhead{Basis of} &
\colhead{Telescope \&} &
\colhead{} \\
\colhead{Name} &
\colhead{Type} &
\colhead{Selection\tablenotemark{a}} &
\colhead{Instrument} &
\colhead{Date}}
\startdata
2MASS J04130959+2440204 &     M5V &    CMDs &  IRTF/SpeX & 2010 Dec 23 \\
2MASS J04155686+2907509 &  galaxy &      IR & Gemini/GMOS & 2009 Oct 20 \\
2MASS J04162497+3037545 &   L1.5V &    CMDs &  IRTF/SpeX & 2010 Jan 3 \\
2MASS J04181059+2844473 &  galaxy &      IR & Gemini/GMOS & 2009 Oct 22 \\
2MASS J04181375+2539395 &     M6V &    CMDs &  IRTF/SpeX & 2010 Jan 4 \\
2MASS J04182560+2836371 &       F &    CMDs &  IRTF/SpeX & 2010 Jan 4 \\
2MASS J04224036+2441144 &   giant &    CMDs &  IRTF/SpeX & 2010 Dec 23 \\
2MASS J04232153+2453513 &   giant &    CMDs &  IRTF/SpeX & 2010 Dec 23 \\
2MASS J04234043+2844151 &     M7V &    CMDs &  IRTF/SpeX & 2010 Dec 23 \\
2MASS J04263650+2439469 &   giant &    CMDs &  IRTF/SpeX & 2010 Dec 23 \\
2MASS J04264776+2456594 &   M4.5V &    CMDs &  IRTF/SpeX & 2010 Jan 4 \\
2MASS J04280370+2424077 &     M5V &    CMDs &  IRTF/SpeX & 2010 Dec 23 \\
WISE J042823.04+264933.8 &  galaxy &      IR & Gemini/NIRI & 2009 Oct 1 \\
2MASS J04305575+2450174 &   M4.5V &    CMDs &  IRTF/SpeX & 2010 Jan 4 \\
WISE J043809.73+254731.5 &  galaxy &      IR & Gemini/NIRI & 2009 Oct 4 \\
2MASS J04381659+2614500 &   giant &    CMDs &  IRTF/SpeX & 2010 Dec 23 \\
2MASS J04435167+2519576 &   M3.5V &    CMDs &  IRTF/SpeX & 2010 Dec 23 \\
2MASS J04511884+2556333 &     M5V &    CMDs &  IRTF/SpeX & 2010 Jan 4 
\enddata
\tablenotetext{a}{Sources were selected for spectroscopy because
they were candidate members based on CMDs or mid-IR excess emission.}
\end{deluxetable}

\begin{deluxetable}{llllll}
\tabletypesize{\scriptsize}
\tablewidth{0pt}
\tablecaption{Spectral Types for Known Members of TWA\label{tab:twa}}
\tablehead{
\colhead{TWA\tablenotemark{a}} &
\colhead{2MASS} &
\colhead{Other Names} &
\colhead{Spectral Type\tablenotemark{b}} &
\colhead{Ref} &
\colhead{Adopt}}
\startdata
  1 & J11015191$-$3442170 &             TW Hya & K7,K8,K6,M2.5(ir),K8,M0.5,M1 & 1,2,3,4,5,6,7 &         M1 \\
  2 & J11091380$-$3001398 &       CD$-$29 8887 & M0.5,M2,M2,M1.5,M1.5,M2.2,M2.25 & 1,2,3,8,5,6,7 &      M2.25 \\
 3A & J11102788$-$3731520 &        Hen 3-600 A & M3,M4,M4,M3.9,M4,M4.1,M4.25 & 1,2,3,8,5,6,7 &      M4.25 \\
 3B & J11102788$-$3731520 &        Hen 3-600 B & M3.5,M4,M3.9,M4,M4.25 &  1,3,8,6,7 &      M4.25 \\
  4 & J11220530$-$2446393 &           HD 98800 &    K5,K7,K5,K6,K6,K7 & 1,2,3,5,6,7 &       K6.5 \\
 5A & J11315526$-$3436272 &       CD$-$33 7795 & M1.5,M3,M2,M1.9,M2,M2.7,M3 & 1,2,3,8,5,6,7 &         M3 \\
 5B &          \nodata &            \nodata &       M8.75,M8.5(ir) &       9,10 &       M8.5 \\
  6 & J10182870$-$3150029 &    Tyc 7183-1477-1 &          K7,M0,M0,M0 &    1,3,6,7 &         M0 \\
  7 & J10423011$-$3340162 &    Tyc 7190-2111-1 &  M1,M2,M3,M3.2,M3.25 &  1,3,5,6,7 &      M3.25 \\
 8A & J11324124$-$2651559 &    GSC 06659-01080 & M2,M2,M3,M2.4,M3,M2.9,M3.25 & 1,2,3,8,5,6,7 &      M3.25 \\
 8B & J11324116$-$2652090 &            \nodata &         M5,M5.2,M5.5 &      1,6,7 &      M5.25 \\
 10 & J12350424$-$4136385 &     GSC 0776-00743 &   M2.5,M2,M2.6,M3.25 &    1,3,8,7 &         M3 \\
11A & J12360103$-$3952102 &          HR 4796 A &                   A0 &          1 &         A0 \\
11B &          \nodata &          HR 4796 B &                 M2.5 &          1 &       M2.5 \\
11C & J12354893$-$3950245 &          HR 4796 C &     M4.5,M4.5,M5(ir) &    11,7,12 &       M4.5 \\
 12 & J11210549$-$3845163 &  RX J1121.1$-$3845 &   M1.5,M1.6,M2,M2.75 &    3,8,5,7 &      M2.75 \\
13A & J11211723$-$3446454 &  RX J1121.3$-$3447 &        M2,M1,M1.1,M1 &   13,3,6,7 &         M1 \\
13B & J11211745$-$3446497 &            \nodata &                   M1 &     13,3,6 &         M1 \\
 16 & J12345629$-$4538075 &     UCAC2 12217020 &      M1.5,M1.8,M2,M3 &   14,8,5,7 &         M3 \\
 20 & J12313807$-$4558593 &    GSC 08231-02642 &          M2,M3,M3.25 &     15,5,7 &      M3.25 \\
 21 & J10131476$-$5230540 &          HD 298936 &          K3/K4,K3,K3 &     16,3,5 &         K3 \\
 23 & J12072738$-$3247002 & SSSPM 1207$-$3247 &         M1,M2.9,M3.5 &     16,8,7 &       M3.5 \\
 25 & J12153072$-$3948426 &     Tyc 7760-283-1 &  M0,M1,K9,M0.5,M0.75 & 16,3,5,6,7 &      M0.75 \\
 26 & J11395113$-$3159214 &            \nodata & M8,M9(ir),M9,M9(ir),M8.5 & 17,18,19,12,7 &       M8.5 \\
27A & J12073346$-$3932539 &            \nodata & M8,M8(ir),M8.25,M8.5(ir),M8(ir),M8.5(ir) & 17,18,20,10,12,21 &      M8.25 \\
27B &          \nodata &            \nodata &               L3(ir) &         12 &      mid L \\
 28 & J11020983$-$3430355 & SSSPM 1102$-$3431 &     M8.5,M9(ir),M8.5 &    22,12,6 &       M8.5 \\
 29 & J12451416$-$4429077 & DENIS 1245$-$4429 & M9.5,M9(ir),M9.25 &    18,18,7 &      M9.25 \\
30A & J11321831$-$3019518 &            \nodata &             M5,M4.75 &       23,7 &      M4.75 \\
30B & J11321822$-$3018316 &            \nodata &                   M4 &         24 &         M4 \\
 32 & J12265135$-$3316124 &            \nodata &           M6.3,M5.25 &        8,7 &      M5.25 \\
 33 & J11393382$-$3040002 &            \nodata &           M4.7,M4.75 &       25,7 &      M4.75 \\
 34 & J10284580$-$2830374 &            \nodata &            M4.9,M5.5 &       25,7 &       M5.5 \\
 35 & J12002750$-$3405371 &            \nodata &                 M4.7 &         26 &       M4.7 \\
 36 & J12023799$-$3328402 &            \nodata &                 M4.8 &         26 &       M4.8 \\
 39 & J10120908$-$3124451 & SCR 1012$-$3124 AB &                   M4 &         27 &         M4 \\
 40 & J12074836$-$3900043 &            \nodata & L0,L1(ir),L1(ir),M9.5(ir) & 28,28,21,7 &      M9.75 \\
 41 & J11472421$-$2040204 & WISEA 1147$-$2040 &               L7(ir) &         29 &         L7 \\
 42 & J11193254$-$1137466 &            \nodata &               L7(ir) &         30 &         L7 \\
 43 & J11084400$-$2804504 &          HIP 54477 &                   A1 &         31 &         A1 
\enddata
\tablenotetext{a}{TWA names 39 through 43 were assigned by \citet{gag16}.}
\tablenotetext{b}{Spectral types measured from near-IR spectra are indicated by
``ir".}
\tablerefs{
(1) \cite{web99};
(2) \cite{tor00};
(3) \cite{tor06};
(4) \cite{vac11};
(5) \cite{pec13};
(6) \cite{her14};
(7) this work;
(8) \cite{shk11};
(9) \cite{neu00};
(10) \cite{bon14};
(11) \cite{kas08};
(12) \cite{all13};
(13) \cite{ste99};
(14) \cite{zuc01};
(15) \cite{rei03};
(16) \cite{zuc04};
(17) \cite{giz02};
(18) \cite{loo07};
(19) \cite{rei08};
(20) \cite{her09};
(21) \cite{gag15};
(22) \cite{sch05};
(23) \cite{loo10a};
(24) \cite{loo10b};
(25) \cite{sch12};
(26) \cite{mur15};
(27) \cite{rie14};
(28) \cite{gag14};
(29) \cite{sch16};
(30) \cite{kel15};
(31) \cite{hou82}.}
\end{deluxetable}

\begin{deluxetable}{lllll}
\tabletypesize{\scriptsize}
\tablewidth{0pt}
\tablecaption{Data for Construction of Standard Spectra\label{tab:stand}}
\tablehead{
\colhead{Name} &
\colhead{Spectral Type} &
\colhead{Ref} &
\colhead{SpeX Date} &
\colhead{Ref}}
\startdata
                  TWA 6 &             M0 &      1 &    2005 Dec 14 &      2 \\
                  TWA 1 &             M1 &      1 &    2005 Dec 14 &      2 \\
                 TWA 13 &             M1 &      1 &     2011 Dec 3 &      2 \\
               UX Tau B &             M2 &      3 &     2008 Nov 2 &      2 \\
                  TWA 2 &          M2.25 &      1 &    2005 Dec 14 &      2 \\
                 TWA 12 &          M2.75 &      1 &    2015 Apr 20 &      2 \\
                DM Tau  &             M3 &      3 &     2008 Nov 2 &      2 \\
                 TWA 10 &             M3 &      1 &    2015 Apr 21 &      2 \\
                  TWA 7 &          M3.25 &      1 &    2005 Dec 14 &      2 \\
                 TWA 8A &          M3.25 &      1 &    2005 Dec 14 &      2 \\
                 LRL 85 &          M3.25 &      4 &    2004 Nov 12 &      6 \\
                 TWA 23 &           M3.5 &      1 &    2015 Apr 19 &      2 \\
                LRL 147 &           M3.5 &      4 &    2004 Nov 12 &      6 \\
                  MHO 9 &          M4.25 &      3 &    2004 Nov 11 &      6 \\
2MASS J04351316+1725496 &           M4.5 &      3 &     2010 Jan 3 &      2 \\
                TWA 11C &           M4.5 &      1 &    2015 Apr 22 &      2 \\
2MASS J04554757+3028077 &          M4.75 &      3 &    2004 Nov 12 &      6 \\
                 TWA 33 &          M4.75 &      1 &    2015 Apr 19 &      2 \\
2MASS J16160602$-$2528217 &          M4.75 &      7 &    2013 Jun 21 &      2 \\
                KPNO 10 &             M5 &      3 &    2004 Nov 11 &      6 \\
              Haro 6-32 &             M5 &      3 &    2004 Nov 12 &      6 \\
2MASS J04555288+3006523 &          M5.25 &      3 &    2004 Nov 12 &      6 \\
                  MHO 7 &          M5.25 &      3 &    2004 Nov 11 &      6 \\
                 TWA 8B &          M5.25 &      1 &    2005 Dec 14 &      2 \\
                 TWA 32 &          M5.25 &      1 &    2015 Apr 21 &      2 \\
                 TWA 34 &           M5.5 &      1 &    2015 Apr 20 &      2 \\
2MASS J16134045$-$2233156 &           M5.5 &      7 &    2012 Apr 25 &      2 \\
2MASS J16002631$-$2259412 &           M5.5 &      7 &    2006 Jun 15 &      2 \\
                 TWA 37 &          M5.75 &      1 &    2015 Apr 20 &      2 \\
                FW Tau  &           M5.8 &      3 &     2008 Nov 2 &      2 \\
                  MHO 8 &             M6 &      3 &    2004 Nov 12 &      6 \\
           V410 X-ray 3 &          M6.25 &      3 &    2004 Nov 11 &      6 \\
2MASS J04552333+3027366 &          M6.25 &      3 &    2004 Nov 12 &      6 \\
2MASS J16051403$-$2406524 &           M6.5 &      7 &    2005 Jun 16 &      2 \\
2MASS J16095852$-$2345186 &           M6.5 &      7 &    2005 Jun 16 &      2 \\
                  MHO 4 &             M7 &      3 &    2004 Nov 11 &      6 \\
                 CFHT 4 &             M7 &      3 &    2004 Nov 11 &      6 \\
               FU Tau A &             M7 &      3 &     2007 Dec 3 &      8 \\
2MASS J16114261$-$2525511 &             M7 &      7 &    2013 Jun 21 &      2 \\
2MASS J04390396+2544264 &          M7.25 &      3 &    2004 Nov 11 &      6 \\
                 KPNO 2 &           M7.5 &      3 &    2004 Nov 11 &      6 \\
                 KPNO 5 &           M7.5 &      3 &    2004 Nov 11 &      6 \\
                 CFHT 3 &          M7.75 &      3 &    2004 Nov 11 &      6 \\
2MASS J04414825+2534304 &          M7.75 &      3 &    2004 Nov 11 &      6 \\
                LRL 405 &             M8 &      5 &    2004 Nov 12 &      6 \\
                 KPNO 7 &          M8.25 &      3 &    2004 Nov 11 &      6 \\
2MASS J04290068+2755033 &          M8.25 &      3 &    2005 Dec 14 &      2 \\
             LH 0429+17 &          M8.25 &      3 &     2007 Dec 3 &      9 \\
                 TWA 27 &          M8.25 &      1 & 2007 Mar 16,2015 Dec 14 &   10,2 \\
                 KPNO 6 &           M8.5 &      3 &    2004 Nov 11 &      6 \\
                 KPNO 1 &           M8.5 &      3 &    2004 Nov 11 &      6 \\
                 TWA 26 &           M8.5 &      1 &    2005 Dec 14 &      2 \\
                 TWA 28 &           M8.5 &      1 &    2005 Dec 14 &      2 \\
2MASS J04263055+2443558 &          M8.75 &      3 &    2004 Nov 13 &     11 \\
2MASS J04334291+2526470 &          M8.75 &      3 &    2005 Dec 13 &     11 \\
                KPNO 12 &             M9 &      3 &    2004 Nov 11 &      6 \\
2MASS J04574903+3015195 &          M9.25 &      3 &    2004 Nov 11 &      6 \\
                 TWA 29 &          M9.25 &      1 &    2007 Mar 17 &     10 \\
                 KPNO 4 &           M9.5 &      3 &    2004 Nov 11 &      6 \\
2MASS J01415823$-$4633574 &             L0 &     12 &     2004 Sep 4 &     13 \\
2MASS J22134491$-$2136079 &             L0 &     12 &     2011 Sep 7 &     14 \\
2MASS J23153135+0617146 &             L0 &     15 &    2007 Nov 13 &     16 \\
2MASS J00550564+0134365 &             L2 &     15 &     2003 Sep 3 &     16 \\
2MASS J05361998$-$1920396 &             L2 &     17 &    2012 Sep 19 &     14 \\
2MASS J05012406$-$0010452 &             L4 &  18,12 &     2010 Jan 3 &      2 \\
2MASS J01262109+1428057 &             L4 &     17 &     2006 Dec 7 &     19 \\
                 TWA 41 &             L7 &      1 &    2016 Feb 11 &     20 \\
                 TWA 42 &             L7 &      1 &   2013 Jun 5-6 &     21 
\enddata
\tablerefs{
(1) Table~\ref{tab:twa};
(2) this work;
(3) Table~\ref{tab:mem};
(4) \cite{luh03ic};
(5) \cite{luh99};
(6) \cite{mue07};
(7) K. Luhman, in preparation;
(8) \cite{luh09fu};
(9) \cite{luh09tau};
(10) \cite{loo07};
(11) \cite{luh06tau1};
(12) \cite{cru09};
(13) \cite{kir06};
(14) \cite{all13};
(12) \cite{cru16};
(16) \cite{fah16};
(17) \cite{fah13};
(18) \cite{rei08};
(19) \cite{met08};
(20) \cite{sch16};
(21) \cite{kel15}.}
\end{deluxetable}

\clearpage

\end{document}